\documentclass[12pt,twoside]{apa}
\usepackage{a4,amssymb,graphicx,eurosym}
\usepackage{apacite}
\usepackage{color}
\usepackage[utf8]{inputenc}
\usepackage[american]{babel}
\usepackage{amsmath}
\usepackage{amsfonts}
\usepackage{amssymb}
\usepackage{csquotes}
\usepackage{gensymb}
\usepackage{fancyhdr}
\usepackage{array}
\usepackage{dcolumn}
\usepackage{booktabs}
\usepackage{lineno}
\usepackage[doublespacing]{setspace}

\makeatletter
\long\def\@makecaption#1#2{%
  \vskip\abovecaptionskip
  \sbox\@tempboxa{#1 #2}%
  \ifdim \wd\@tempboxa >\hsize
    #1 #2\par
  \else
    \global \@minipagefalse
    \hb@xt@\hsize{\hfil\box\@tempboxa\hfil}%
  \fi
  \vskip\belowcaptionskip}
\makeatother

\begin{document}

\thispagestyle{empty}
\pagenumbering{roman}
\begin{center}
{\huge\bf
Influence of initial fixation position \\[0.3ex] 
in scene viewing}\\ 
\vspace{10mm}
\large
Lars O.~M.~Rothkegel$^{1*}$, Hans A.~Trukenbrod$^{1}$, Heiko H.Sch\"utt$^{1,2}$, \\[1ex]
Felix A.~Wichmann$^{2-4}$, and Ralf Engbert$^{1}$
\\
\vspace{10mm}
\normalsize
$^1$University of Potsdam, Germany\\
$^2$Eberhard Karls Universit\"at T\"ubingen, Germany\\
$^3$Bernstein Center for Computational Neuroscience T\"ubingen, Germany\\
$^4$Max Planck Institute for Intelligent Systems, T\"ubingen, Germany\\
\vspace{10mm}
\today
\end{center}

\vspace{10mm}
\noindent
\vspace{\fill}

\noindent
$^*$To whom correspondence should be addressed: \\
Lars Rothkegel \\
Department of Psychology \& Cognitive Science Program  \\ 
University of Potsdam\\ 
Am Neuen Palais 10\\ 
14469 Potsdam\\
Germany \\ 
E-mail: lars.rothkegel\symbol{64}uni-potsdam.de \\
Phone: +49 331 9772370, Fax: +49 331 9772794

\newpage
\ 
\newpage
\setcounter{page}{1}
\pagenumbering{arabic}
\pagestyle{myheadings}
\markboth{Rothkegel et al.}{Influence of initial fixation position}

\doublespacing
\section*{Abstract} 
During scene perception our eyes generate complex sequences of fixations. Predictors of fixation locations are bottom-up factors like luminance contrast, top-down factors like viewing instruction, and systematic biases like the tendency to place fixations near the center of an image. However, comparatively little is known about the dynamics of scanpaths after experimental manipulation of specific fixation locations. Here we investigate the influence of initial fixation position on subsequent eye-movement behavior on an image. We presented 64 colored photographs to participants who started their scanpaths from one of two experimentally controlled positions in the right or left part of an image. Additionally, we computed the images' saliency maps and classified them as balanced images or images with high saliency values on either the left or right side of a picture. As a result of the starting point manipulation, we found long transients of mean fixation position and a tendency to overshoot to the image side opposite to the starting position. Possible mechanisms for the generation of this overshoot were investigated using numerical simulations of statistical and dynamical models. We conclude that inhibitory tagging is a viable mechanism for dynamical planning of scanpaths.

\section{Introduction}
An important problem for research on human vision is to predict where people look in visual scenes \cite{tatler2008systematic}. Recording of eye movements is among the most important tools to investigate how attention is distributed over a give scene \cite{findlayactive}. In addition to scene content \cite{henderson2003human}, image-independent viewing strategies seem to exist, e.g., a central fixation tendency \cite{tatler2007central} as the most important effect in this category. To obtain a deeper understanding about dynamical aspects of the attention distribution over a scene and possible dependencies between successive fixations we investigate the influence of the eye's starting position on subsequent viewing behavior based on statistical and dynamical assumptions about eye guidance.

Processes that influence the selection of upcoming saccade targets can be divided into three different categories of theoretical principles that are typically discussed. {\em Bottom-up processes} derive from properties of the viewed stimulus \cite{mannan1996relationship,itti1998model,parkhurst2002modeling}. {\em Top-down processes} depend on the mental state of an observer, e.g., the observers' visual memory \cite{henderson2003eye} or the instruction given to the observer before inspection of a scene \cite{yarbus1967eye,castelhano2009viewing}. Finally, {\sl systematic tendencies} describe eye movement behavior found in many experiments independent of stimulus and observer. The initial selection of the center of an image \cite{tatler2007central,bindemann2010scene}, the tendency to make initial movements in the leftward direction \cite{dickinson2009spatial,foulsham2013leftward,ossandon2014spatial} or the preference for horizontal over vertical saccades \cite{foulsham2010asymmetries} belong to this category.

Research on bottom-up processes has been particularly successful to predict fixation locations from low-level image features such as contrast, orientation and color \cite{itti1998model,torralba2003modeling,kienzle2006nonparametric}. For a given scene, computational models generate a {\sl saliency map}, a 2D probability distribution that indicates the probability of receiving a fixation in an eye tracking experiment with human participants \cite{itti1998model,itti2000saliency,judd2012benchmark,borji2013state}. Thus, a saliency map is a stationary model that computes probabilities for all locations simultaneously.

However, current computational models for the prediction of fixation locations are not exclusively based on bottom-up features. Although the original meaning of {\em saliency} refers to the bottom-up features of an image, newer computational models that include other features are also termed saliency models by their authors \cite{judd2009learning,mit-saliency-benchmark}. Therefore, we will refer to all stationary models that aim at the prediction of fixation locations as saliency models.  Recent models incorporate top-down processes like the task demands \cite{navalpakkam2005modeling} and other higher-level image features like face processing \cite{cerf2008predicting}. Moreover, systematic tendencies such as the central fixation bias \cite{tatler2007central} are included in the computation of saliency maps. As a result, current models integrate multiple features from all three categories of processes into a coherent computational framework \cite{cerf2008predicting,judd2009learning}. 

All saliency models need to predict the density of the eye's fixation locations (so-called first-order statistics). Thus, the evaluation of saliency models is primarily based on the assumption of statistically independent fixations without reference to previous fixations (i.e., the scanpath). Compared to static saliency models, dynamic models try to capture some additional aspects of the scanpath. Dynamical principles for saccade planning are {\em inhibitory tagging} \cite{klein1988inhibitory,itti1998model,bays2012active,le2015saccadic}, {\em saccadic momentum} \cite{smith2009facilitation,smith2011does,wilming2013saccadic} and {\em facilitation of return} \cite{smith2009facilitation,smith2011does,luke2013temporal}

Inhibitory tagging is motivated by the effect of inhibition of return, a neural mechanism that inhibits the processing at recently attended locations \cite{posner1984components,posner1985inhibition,klein2000inhibition} and is often interpreted as a foraging facilitator. While this mechanism was first discovered as an effect on a temporal scale, i.e., increased processing time at a previously attended stimulus for a specific time window, inhibition of return might carry over to spatial effects. In the case of spatial inhibition of return recently fixated positions are inhibited from being re-fixated shortly afterwards \cite{gilchrist2000refixation}. Several studies were unable to report evidence for inhibition of return during scene viewing; quite the contrary, a facilitation of return saccades to currently fixated locations has been found \cite{smith2009facilitation,smith2011does,wilming2013saccadic}. 

However, compared to statistical baseline model without memory based on inhibitory tagging, return saccades occur less often in experiments than expected \cite{bays2012active}, when the density map of fixations and the distribution of angles between two subsequent saccades are reproduced. Therefore, there is at least weak support for a memory-producing mechanism during scene exploration. In agreement with this result, we recently published a computational model of saccade generation in scene viewing that implemented both inhibitory tagging and dynamical attention mechanisms \cite{engbert2015spatial}. In model inhibitory tagging is combined with a dynamical activation map representing attention allocation, allowing the model to reproduce second-order statistics that include spatial correlation functions characterizing the clustering of fixations in addition to the first-order density of fixations. Thus, inhibitory tagging seems to be important to reproduce higher-order scanpath statistics \cite{engbert2015spatial}, despite the current lack of direct experimental support for inhibition of return in scene viewing \cite{smith2009facilitation,smith2011does,luke2013temporal}.

Saccadic momentum, another dynamical principle of saccade planning in scene viewing, describes the tendency to maintain the direction of the previous saccade for the upcoming saccade \cite{smith2009facilitation,smith2011does,wilming2013saccadic}. Similar to inhibition of return, saccadic momentum could serve as a foraging facilitator in visual search. Finally, {\em facilitation of return} describes the tendency that it is actually more likely to produce return saccades than it would be by chance 
\cite{hooge2005inhibition,smith2009facilitation}. On the time scale of one fixation duration ($\sim 300$~ms), such a facilitation seems to be in contradiction to spatial inhibitory tagging. Because of these behaviorally relevant ongoing neurocognitive processes, we were interested to find experimental support for the presence of {\em inhibitory tagging}, {\em saccadic momentum}, {\em facilitation of return} or a mixture of these fundamental principles in attentional and oculomotor control. 

Smith and Henderson (2009) ruled out inhibitory tagging, since they found an increased number of return saccades in comparison to a probabilistic baseline \cite{smith2009facilitation}. However, it has also been argued that there is a reduced number of return saccades compared to a memoryless system \cite{bays2012active}. Given the current mixed evidence on return saccades, we focus on the time window of events. Return saccades are limited to a time window of one fixation duration, i.e., about 300~ms. Since attention moves to the future fixation location before a saccade is executed \cite{deubel1996saccade}, inhibition of return is at its maximum if we assume that the typical time-course transfers to scene viewing \cite{posner1984components}. However, first, it would not be surprising to that more time than a single fixation duration is needed to build-up spatial inhibition. Second, return saccades might be planned before the inhibition of return mechanism is activated, so that saccades to previously inspected image regions could be produced while inhibition is on the rise. Third, it has been reported that the time scale of IOR  is dependent on task difficulty \cite{klein2000inhibition}. Therefore, the current lack of direct evidence for inhibition of return does not rule out inhibitory tagging as a saccade-planning mechanism.

To investigate inhibitory tagging, saccadic momentum, and facilitation of return, we recorded observers' scanpaths on natural scenes starting from one of two predefined starting positions close to either side of the monitor. Participants were forced to maintain fixation at an initial location in an image for one second under gaze-contingent monitoring. Under the hypothesis that spatial inhibitory tagging is active at the starting position, we expected observers (i) to leave their starting positions when fixation markers disappeared, and (ii) not to return immediately to the region of the experimentally controlled starting position.  Since we hypothesized that both behaviors  depend on the saliency of the region of the starting position, we classified natural images into three categories with left-sided and right-sided saliency asymmetry as well as images with an approximately symmetrical saliency distribution. We expected that initial fixations stay closer to the starting position when the starting position was in the more salient region of a scene; alternatively, gaze was expected to move immediately to the opposite side of a scene, when the starting position was in the less salient region of the scene. According to the saccadic momentum and facilitation of return hypothesis, we expected a behavior where subsequent eye movements depend on the direction of the first saccade. With the typical center bias we assume that the gaze had to shift to the center and, subsequently, either maintain direction and move to the opposite image side (saccadic momentum) or return close to the starting position (facilitation of return). 

Below we report scanpath patterns where gaze positions of the participants moved further away from the starting position than predicted by the empirical saliency map or a saccadic momentum mechanism. Next, we compare experimental data with numerical simulations from a range of stochastic models, a model reproducing the saccadic momentum mechanism and dynamical model (SceneWalk) which uses inhibitory tagging as a mechanism for saccade planning \cite{engbert2015spatial}.

\section{Method}
\subsection{Experiment}
\subsubsection{Stimuli}
\begin{figure}[t]
\unitlength1mm
\begin{picture}(150,115)
\put(4.5,-2){\includegraphics[width=150mm]{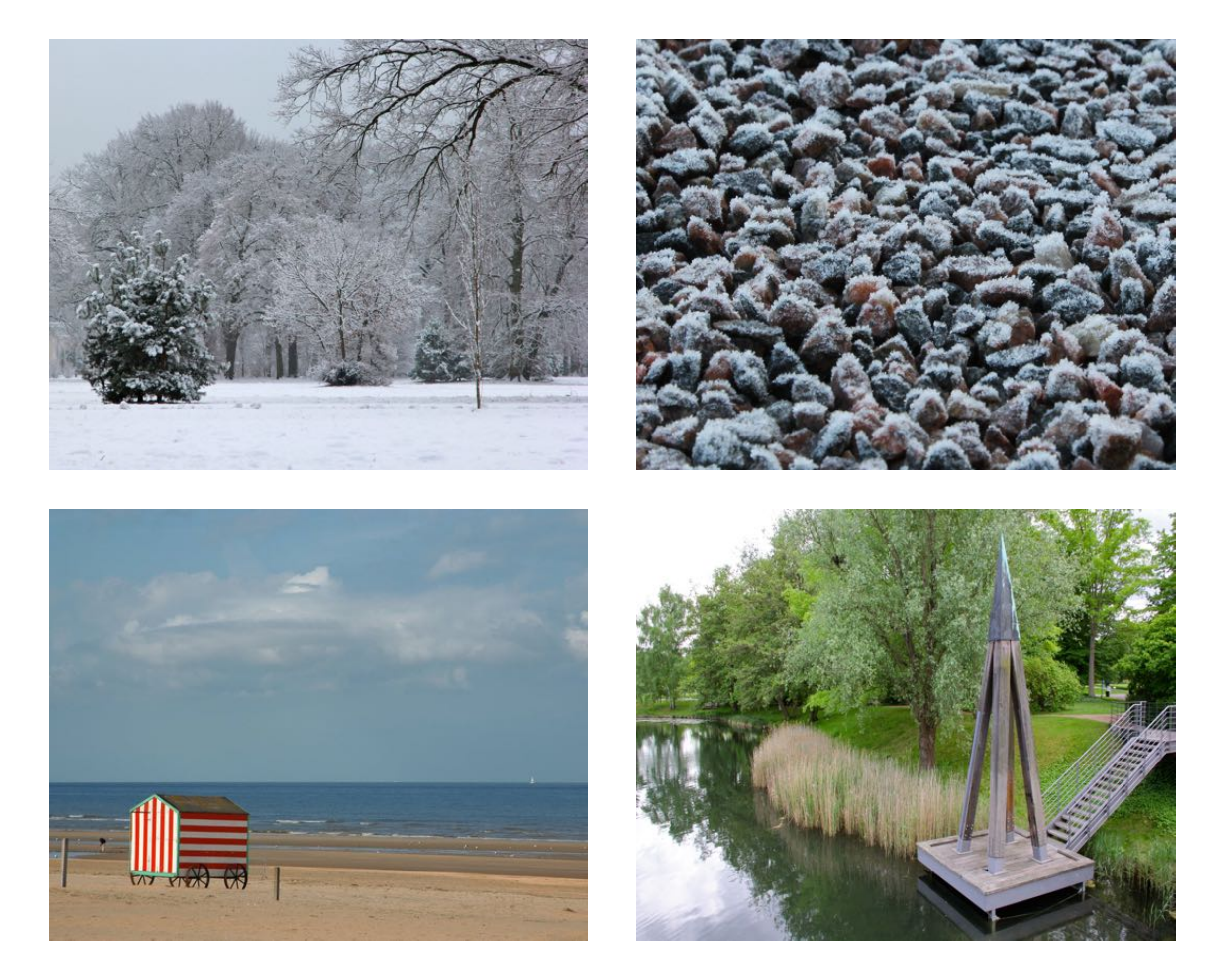}}
\put(4,62){(a)}
\put(77,62){(b)}
\put(4,4){(c)}
\put(77,4){(d)}
\end{picture}
\caption{\label{FigImages}
Examples from the set of images, (a) balanced (b) natural pattern (c) left focus (d) right focus.}
\end{figure}
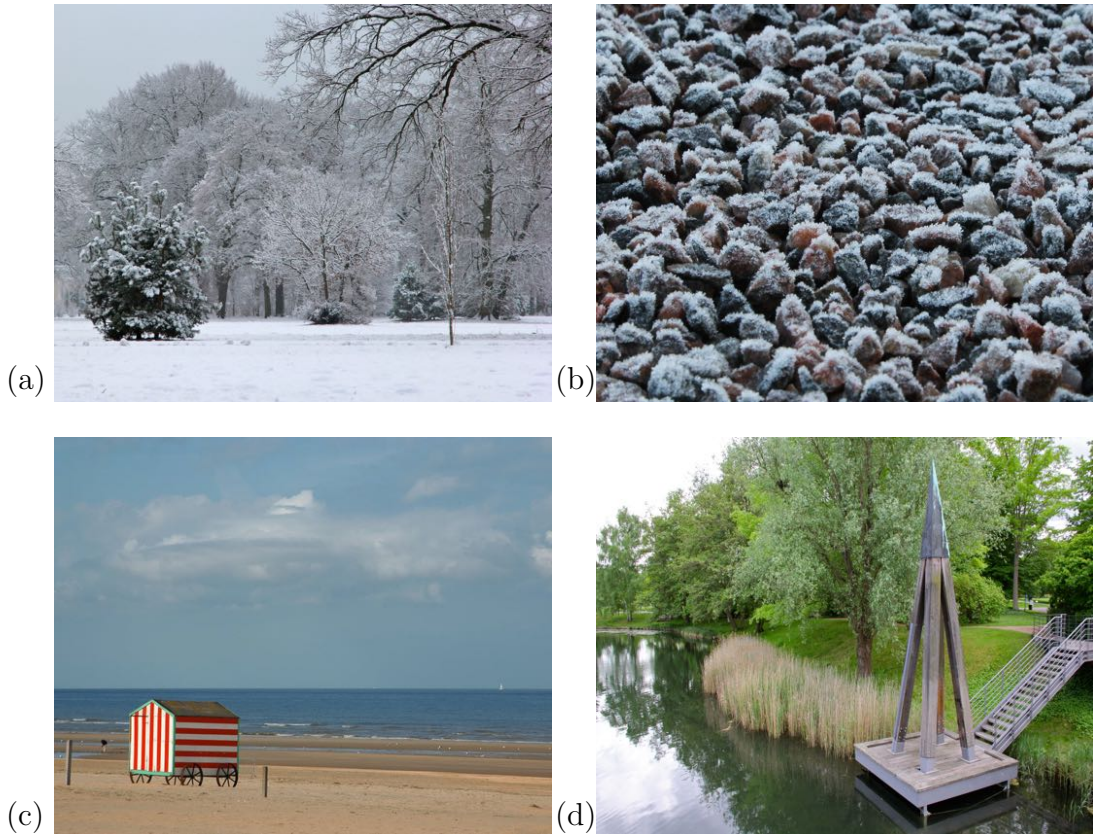

A set of 64 color photographs was presented to human observers. Pictures were presented on a 20 inch CRT monitor (Mitsubishi Diamond Pro 2070; frame rate 120~Hz, resolution 1280~$\times$~1024 pixels; Mitsubishi Electric Corporation, Tokyo, Japan). Images showed either natural object-based scenes (48) or abstract natural patterns (16). Object-based scenes were further devided into three categories as balanced, left focus, or right focus, yielding a total of 4 categories (Fig.~\ref{FigImages}). The Pattern images were chosen to obtain a more homogenous fixation distribution because of the lack of objects present. Systematic oculomotor biases were expected to be stronger in these images. 

\begin{figure}[h]
\unitlength1mm
\begin{picture}(150,80)
\put(2,-2){\includegraphics[width=16cm,height=8cm]{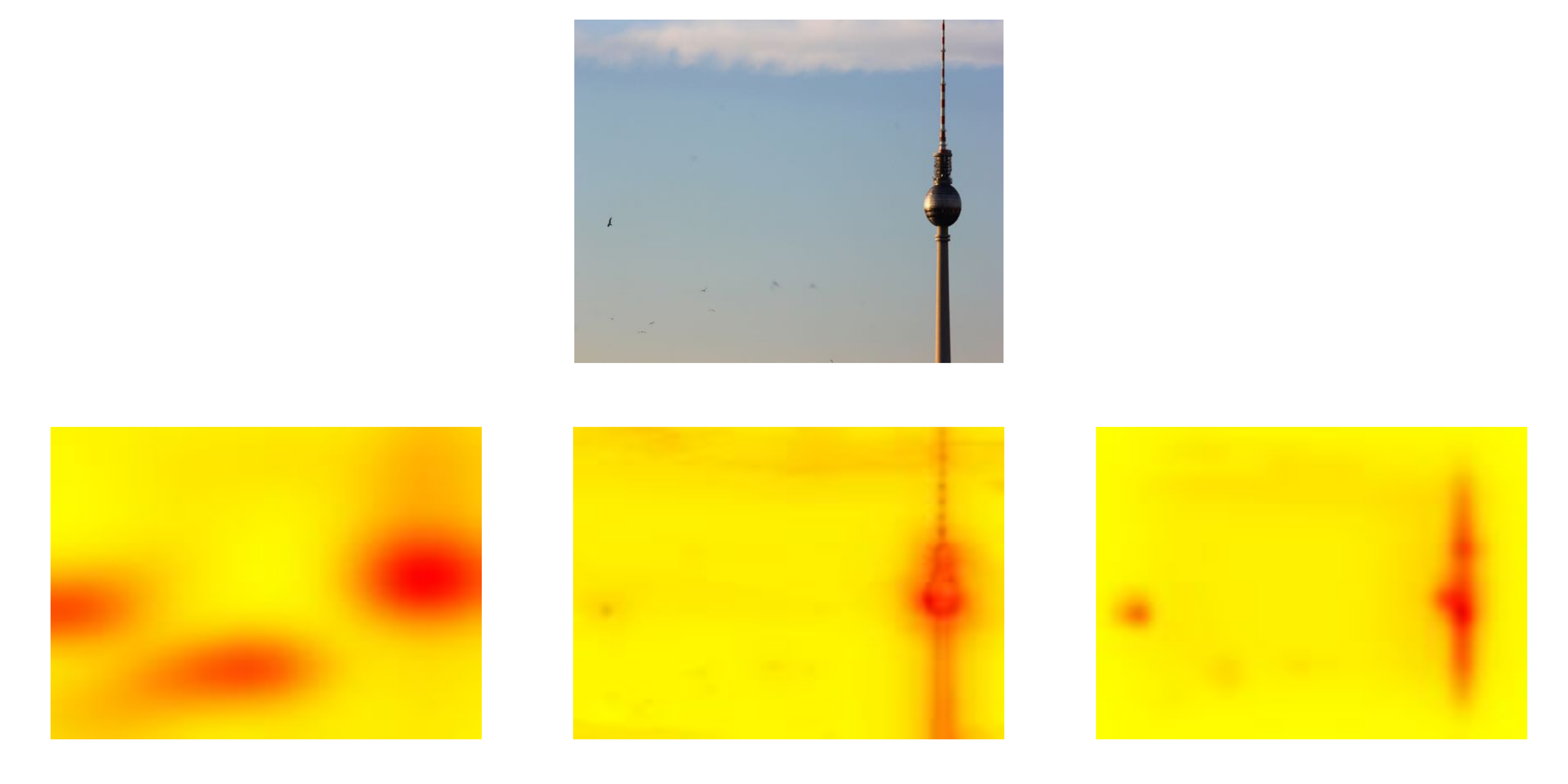}}
\put(54,42){(a)}
\put(0,3){(b)}
\put(54,3){(c)}
\put(107,3){(d)}
\end{picture}
\caption{\label{FigSaliency}
Objective categorizing of images. (a) Example of an image with right focus. (b) Experimental density map of fixations, estimated using a Gaussian kernel with bandwith $\sigma=2.56^\circ$ according to Scott's rule. (c) Output from the Judd saliency model, without distance-to-center analysis and face/object detection. (d) Output from the graph based visual saliency (GBVS) model.}
\end{figure}

For the categorization we used an objective test by computing visual saliency by the graph based visual saliency model \cite{harel2006graph} and the Judd saliency model \cite{judd2009learning} without distance to center weighting and face or object detection. As posthoc measure, the density map of the observers' fixations for each of the 48 natural scenes  was evaluated so that an empirical measure of left and right bias of the images was included. Figure \ref{FigSaliency} shows an example of an image with right focus compared to the output of the two saliency models and the kernel density estimate of the fixation density (excluding the initial fixation) from all observers. To obtain a quantitative measure for the presence of a left, or right focus, we computed the horizontal position of a vertical line, so that there was equal saliency/intensity on each side of the line. If the horizontal position of this line differed by more than 5 percent from the center (for the average over the two saliency models and the human fixation map), the corresponding image was classified as having left or right focus. After application of this criterion, we retained 23 images with focus close to the center (balanced images), 12 images with left focus, and 13 images with right focus among the set of object-based scenes \footnote[1]{We originally chose the natural scenes to contain 16 images in each category. Because our subjective categorization did not match the objective criterion for 7 of the images, an unequal number of images in each category remained for further analysis.}. For the presentation during the experiment, images were converted to a size of 1200~$\times$~960 pixels and displayed in the center of the screen with gray borders extending 32 pixels to the top/bottom and 40 pixels to the left/right of the image. The image covered 31\degree \ of visual angle in the horizontal and 25\degree \ in the vertical dimension. 

\subsubsection{Participants}
We recorded eye movements from 28 human participants with normal or corrected-to-normal vision. The group of participants consisted of 20 female and 8 male observers aged between 19 and 33 years; all were recruited from the University of Potsdam.  Participants received credit points or a monetary compensation of 8~Euro for their participation. The work was carried out in accordance with the Declaration of Helsinki. Informed consent was obtained for experimentation by all participants.

\subsubsection{Procedure}
\begin{figure}
\unitlength1mm
\begin{picture}(150,73)
\put(0,-6){\includegraphics[width=12cm]{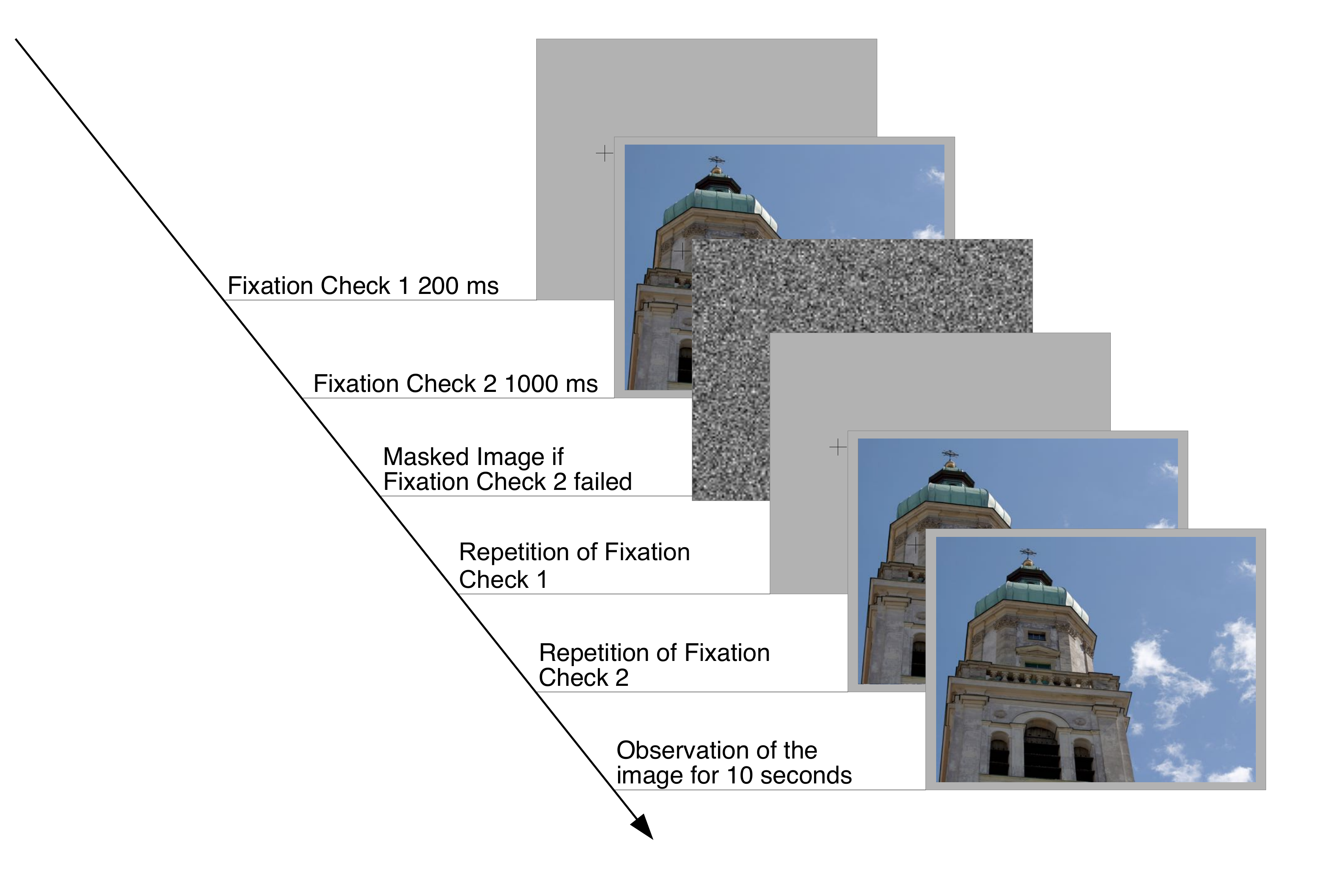}}
\end{picture}
\caption{\label{FigProcedure}
Schematic illustration of the experimental procedure. In the example, the first fixation check failed. After fixation check 2, exploration of the image started.}
\end{figure}

Participants were instructed to position their heads on a chin rest in front of a computer screen at a viewing distance of 70~cm. Eye movements were recorded binocularly using an Eyelink 1000 video-based eye tracking system (SR Research, Osgoode/ON, Canada) with a sampling rate of 1000~Hz. Trials began with a black fixation cross presented on a grey background at the vertical meridian 5.6\degree \ away from the left or right border of the monitor. After successful binocular fixation in a quadratic area with a length of 2.2\degree \  an image appeared while the fixation cross remained present for another second. Participants were instructed to keep their eyes on the fixation cross until it disappeared. This was done to assure that participants started their exploration from the experimentally controlled position. If this fixation test failed, a mask with random noise appeared and the fixation check was repeated. After successful completion of the fixation test participants explored each scene for a subsequent memory test. In the memory test participants had to indicate for 64 Images---32 already presented images and 32 new images---if they had seen it before.\footnote[2]{Participants answered correctly in 91.5\% of the trials with a mean reaction time of 1.4 seconds.}  Figure \ref{FigProcedure} summarises the experimental procedure. In the example, the first fixation check failed, before the actual scene exploration started. Because a fixation check of 1 second was very difficult for participants it had to be repeated in 44\% of all trials. All analyses were conducted separately for the trials with and without a repetition of the second fixation check and there were no systematic differences. Thus for the analyses in this article we used fixations from all trials.

\subsection{Data Analysis}
\subsubsection{Data preprocessing and saccade detection}
For saccade detection we applied a velocity-based algorithm \cite{engbert2003microsaccades,engbert2006microsaccades}. Saccades had a minimum amplitude of $0.58\degree$ and exceeded the average velocity during a trial by 6 standard (median-based) deviations for at least 6 data samples (6~ms). The epoch between two subsequent saccades was defined as a fixation. The number of fixations for further analyses was $47\,330$.

\subsubsection{Mean horizontal distance from starting position}
To analyse the potential dependence of the scanpath on the experimentally controlled starting position, we estimated the temporal evolution of the mean horizontal gaze position. In the first step, we computed the absolute value of the time-dependent horizontal distance to the starting position for each trial. The calculation was based on fixation positions and fixation durations obtained from data preprocessing. The estimated mean horizontal distance (MHD) from starting position was computed as 
\begin{equation}
\label{Eq_MHD}
X_{\rm MHD}(t) = \frac{1}{m\cdot n}\sum_{j=1}^n \sum_{k=1}^m (x_{jk}(t)-x_{jk}(0)) \;,
\end{equation}
where $x_{jk}(t)$ indicates horizontal gaze position at time step $t$ (in milliseconds) for participant $j$ and image $k$. For each combination of image and participant, the starting position $x_{jk}(0)$ was either left or right (see Procedure). To obtain a comparable measurement for both starting positions, gaze position for right starters was mirrored on the vertical central line. Afterwards a Gaussian kernel with $\sigma=100$~ms was applied to obtain a smoothed curve of $\bar{X}_{\rm MHD}(t)$. Another possible analysis would be the vertical or the overall distance to the starting position. The vertical distance showed no interesting effect, as the starting position was always on the vertical midline. The overall distance did hence only depend on the horizontal distance, which was, therefore, our analysis of choice.

\subsection{Model simulations with controlled initial positions}
To interpret the experimental results of the temporal evolution of mean horizontal distance $X_{\rm MHD}(t)$ we performed numerical simulations using statistical control models, a model emulating saccadic momentum and a recently proposed dynamical model for scanpath generation using inhibitory tagging \cite{engbert2015spatial}. For the model runs, simulations started at initial positions corresponding to the experimentally manipulated starting positions. Fixation durations and number of fixations in each trial were equal to the experimental data.  We obtained the same number of trials from numerical simulations as from the experimental data und performed the same analysis on the MHD function $X_{\rm MHD}(t)$.

\subsubsection{Sampling from density map}
As the most straightforward statistical control, we simulated scanpaths by randomly sampling from the 2D density map of all fixations (or empirical saliency map) generated by all participants for a given image. First, we applied kernel density estimation using the SpatStat package \cite{SpatStat} of the R Language of Statistical Computing \cite{R}. Based on a Gaussian kernel function with a bandwidth parameter according to Scott's rule \cite{scott2012multivariate}, ranging from 1.81\degree \ to 2.72\degree, we computed the fixation density map for each image. Second, to simulate a scanpath (i.e., a fixation sequence), we sampled randomly from this map where local density at a particular location translated into probability to generate a fixation at this position. Fixation durations and numbers of fixations were chosen as in the corresponding experimental data  

\subsubsection{Gaussian Model}
Next, we implemented a statistical model that sampled from the empirical saliency map via a Gaussian-shaped aperture to mimic a limited attentional span for saccade target selection. For a given fixation position $x$, the empirical saliency map was weighted by a two dimensional Gaussian, centered at $x$, with a standard deviation of $4.88\degree$ visual angle. This is the same standard deviation used for the attention map of the SceneWalk model by Engbert et al. (see after next section). Sampling from the resulting weighted map, which was recomputed after each fixation, generated a scanpath in this model. As in the previous model, simulations were run with experimentally observed fixation durations, numbers of fixations and starting positions. Effectively, this model is similar to the SceneWalk model without an inhibitory tagging mechanism.

\subsubsection{Saccadic Momentum Model}
The third model reproduced the behavior that saccades, on average, tend to follow the direction of the previous saccade---a phenomenon termed saccadic momentum \cite{smith2009facilitation}. This was applied in a similar way as in a recently published model of saccade generation \cite{le2015saccadic}. In order to reproduce the typical angles between two subsequent saccades, while keeping the saccade amplitude distribution similar to the experimental data, saccades were sampled from the joint probability distribution of amplitudes and angles. To also include the behavior produced by the manipulation of the starting position, the first two saccades of each trial were taken from the experimental data. Again, numbers of fixations and fixation durations were also taken from the experimental data.

\subsubsection{Simulations of the SceneWalk model}
In a recently proposed mathematical model of scanpath generation in scene viewing \cite{engbert2015spatial}, it was assumed that eye movements are driven by the interaction of two neural activation maps. A fixation map $f(x;t)$ keeps track of previous fixations by adding activation at fixation position $x$. The time dependence of this map results from the addition of activation at each time step in combination with fixation-position independent decay. Thus the fixation map implements {\sl inhibitory tagging} \cite{itti2001computational}. The distribution of visual attention at time $t$ is given by a second activation map  $a(x;t)$. The assumption of maps of visual space is consistent with recent neurophysiological work on an allocentric motor map in the primate entorhinal cortex \cite{killian2012map,stensola2012entorhinal}, which is spatially discrete like that in the model with discrete activations $f_{ij}(t)$ and $a_{ij}(t)$, where subscripts $i$ and $j$ denote horizontal and vertical dimensions. 

In the SceneWalk model, the difference of the normalized fixation map $f_{ij}(t)$ and the normalized attention map $a_{ij}(t)$ is a time-dependent potential function $u_{ij}(t)$ computed as  
\begin{equation}\label{Eq_Pot}
u_{ij}(t) = -\frac{a_{ij}(t)}{\sum_{kl}a_{kl}(t)} 
+ \frac{[f_{ij}(t)]^{\gamma}}{\sum_{kl}[f_{kl}(t)]^{\gamma}}  \;,
\end{equation}
where the exponent $\gamma$ is a free parameter that is important for controlling the amount of aggregation (or clustering) of realized gaze positions \cite{engbert2015spatial}. 

Since the potential $u_{ij}(t)$ is the difference of activation maps, it can be positive or negative at position $(i,j)$.  We implemented stochastic selection of saccade targets proportional to relative activations \cite{luce1959individual} among the lattice sites with negative values $(S)$. The probability for saccadic target selection is given by
\begin{equation}\label{Eq_saccsel}
\pi_{ij}(t) = \max_{(k,l)\in {\cal S}}\left(\frac{u_{ij}(t)}{\sum_{kl} u_{k,l}(t)},\,\eta\right) \;,
\end{equation}
where the noise term $\eta$ is an additional model parameter. Since the model predicts fixation locations, but not fixation durations, we used the exact fixation durations from the experimental data for our simulations. Also numbers of fixations and starting positions were chosen as in the experimental data. All model parameters were chosen as in the published version of the SceneWalk model \cite{engbert2015spatial}. In an additional model run, the parameter controlling the strength of the inhibition map was manually adapted for a second analysis with reduced influence of inhibitory tagging.

\section{Results}
In our experiment, we manipulated starting positions to investigate the influence on scanpath statistics. We begin with reporting summary statistics on saccade amplitudes and saccade turning angles, before we analyze the temporal evolution of the mean horizontal distance from the starting position. The temporal evolution of the mean horizontal distance from the starting position will turn out to be an important measure of scanpath statistics. Finally, we run several numerical model simulations to interpret potential mechanisms underlying scanpath generation.

\subsection{Saccadic Amplitudes and Directions} 
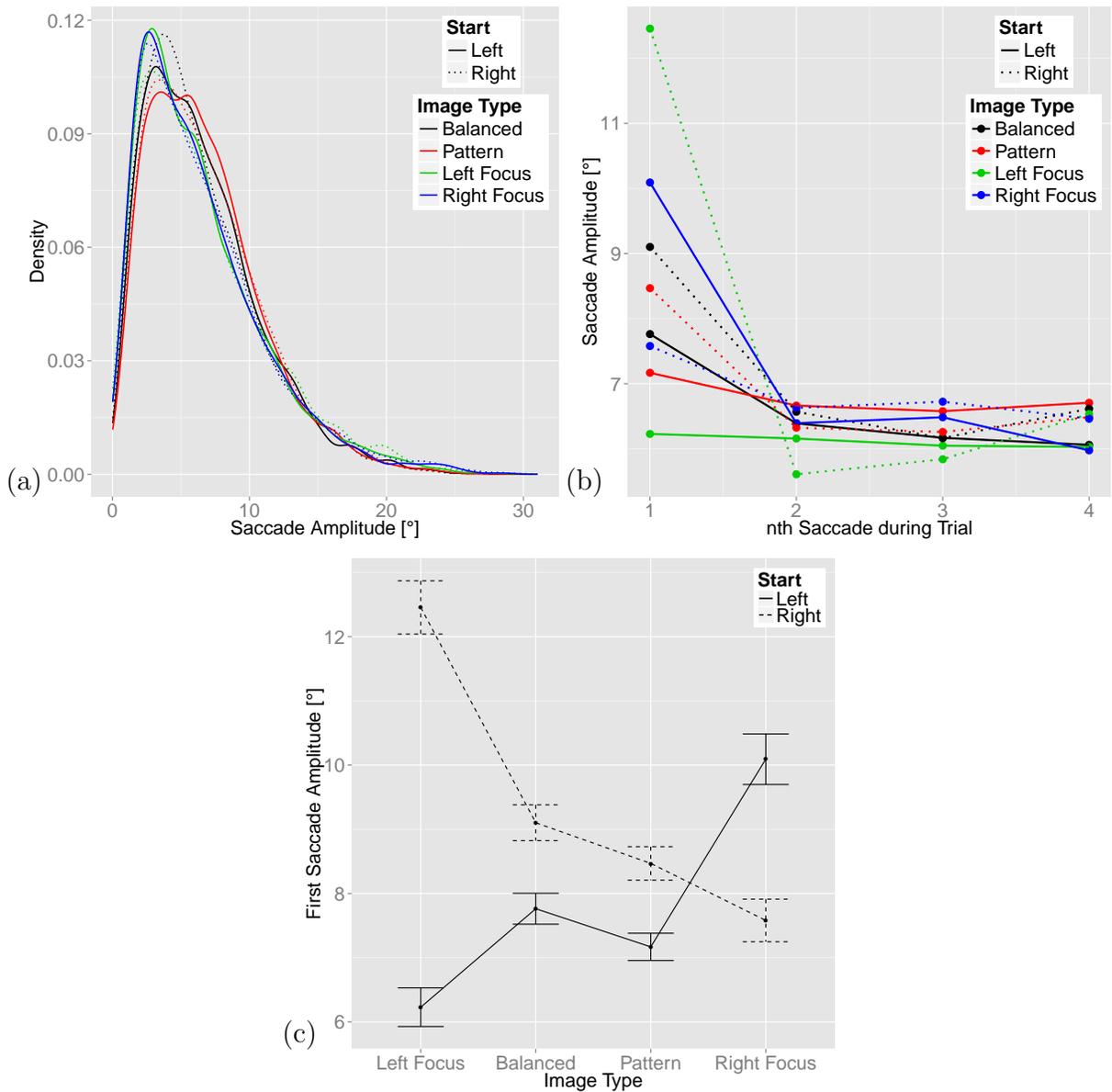
\begin{figure}
\unitlength1mm
\begin{picture}(120,160)
\put(2,80){\includegraphics[width=16cm]{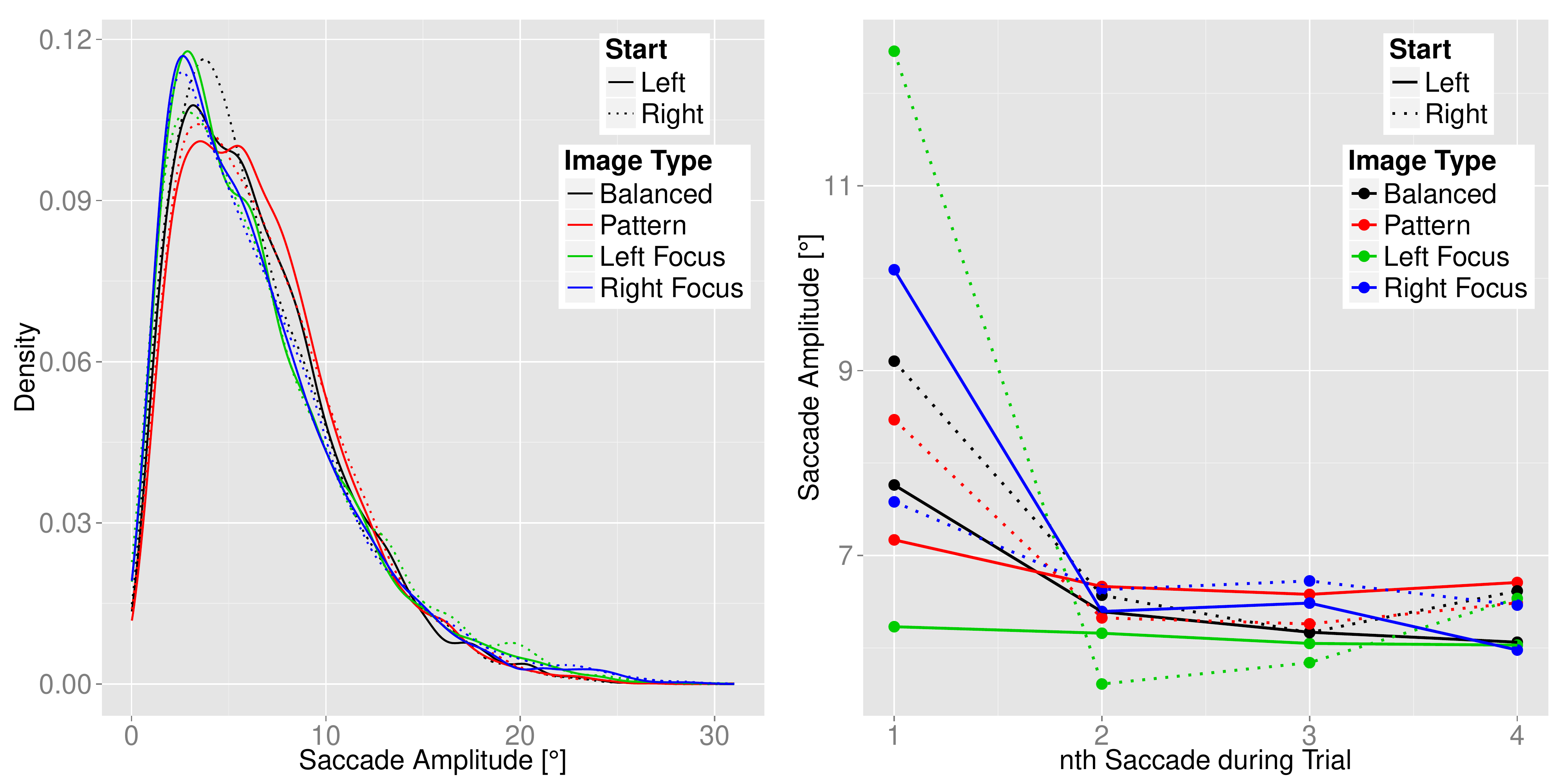}}
\put(0,88){(a)}
\put(81,88){(b)}
\put(42,0){\includegraphics[width=8cm]{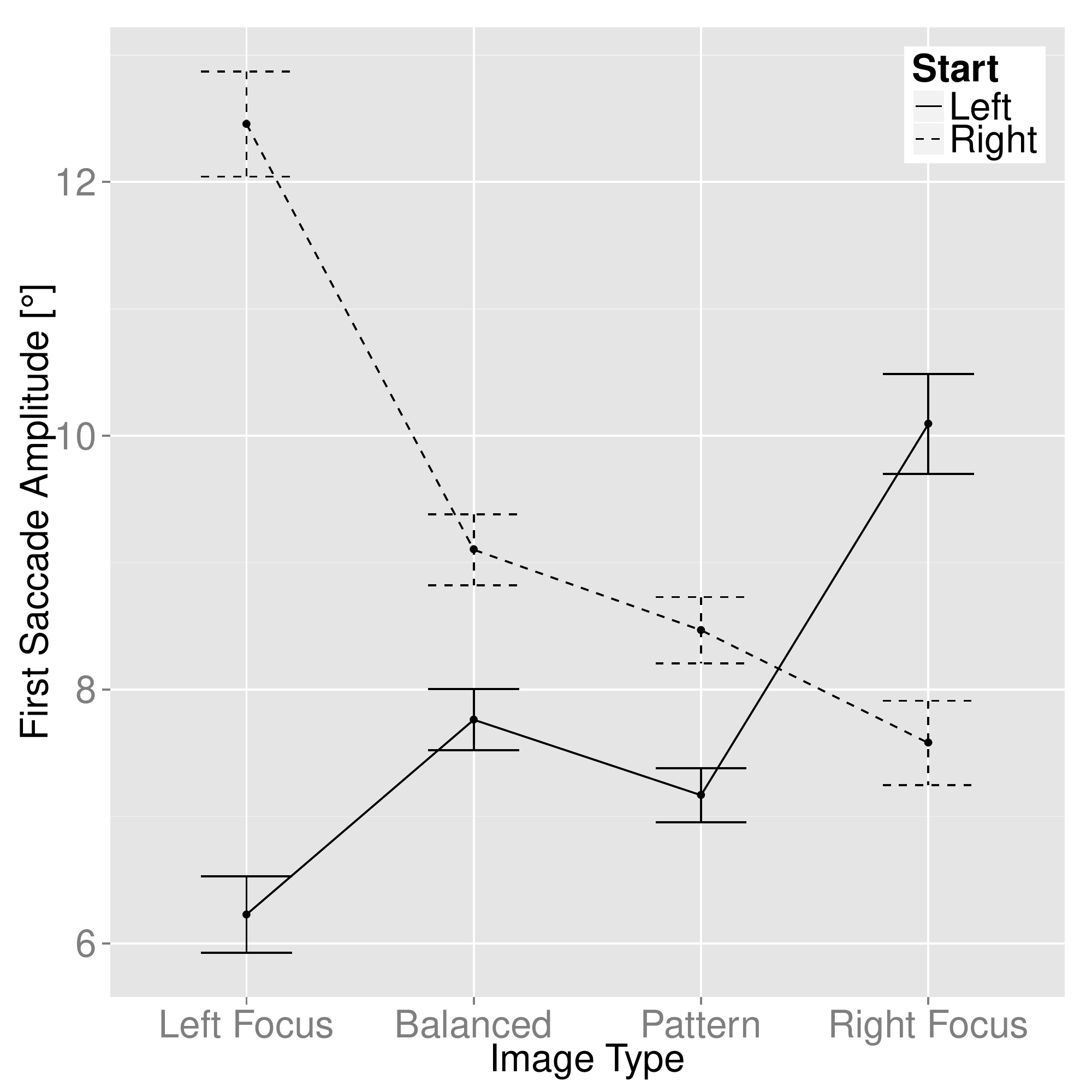}}
\put(40,8){(c)}
\end{picture}
\caption{\label{FigSaccAmp}
Summary statistics of saccade amplitudes. (a) Densities of all saccade amplitudes for the three images types of object-based scenes (balanced, left, and right focus) and the pattern images for left and right starting position. (b) Mean saccade amplitude for the $n$th saccade in each trial for all conditions. While there is a strong effect on the mean of the first saccade length, subsequent saccade amplitudes show no systematic pattern. (c) Mean values of the first saccade amplitude for the 8 different conditions. There is a strong interaction between the image type and the starting position especially for the left focus images and the right focus images. Errorbars represent the standard error of the mean.}
\end{figure}

In our experiment, distributions of saccade amplitudes show the heavy tailed curve that is typically observed in scene viewing experiments \cite{tatler2006long,henderson1998eye}. Saccade amplitude distributions (Fig.~\ref{FigSaccAmp}a) across different image types and starting positions were all very similar. The only visible difference was a slight shift from short to medium saccade lengths in the pattern images compared to the object based images. 
 
The mean amplitude of the first saccade in each trial differed significantly between combinations of the factor starting position and image type. The mean first saccade amplitudes for left and right starters were $\bar{s}_{\rm left}=7.80\degree$ and $\bar{s}_{\rm right}=9.26\degree$, resp. We computed a linear mixed-effects model using the lme4 package \cite{bates2013lme4} in the R Language of Statistical Computing \cite{R} as
\begin{equation}\label{Eq_Model1} 
model= saccade length \sim imagetype  \times  starting position + (1| Subject) + (1|Image)
\end{equation} with the first saccade amplitude as the dependent variable, the starting position, image type and their interaction as fixed effects and the intercept of the subjects and images as random effects.  By using the lmertest package \cite{kuznetsova2013lmertest} and computing the satterthwaite estimation \cite{satterthwaite1946approximate} we obtained $p$-values.  The starting position was significant $(p =1.72 \times 10^{-8})$ as well as image type $(p=9.87 \times 10^{-4})$. Mean values were $ \bar{s}_{balanced}= 8.43 \degree $ , $ \bar{s}_{pattern}= 7.81 \degree $ , $ \bar{s}_{left focus}= 9.34 \degree $ and $ \bar{s}_{right focus}= 8.83 \degree $.  The interaction between image type and starting position was also significant $(p= 4.97 \times 10^{-3})$. Figure \ref{FigSaccAmp}c visualizes this interaction and the main effects of image type and starting position. Computing an ANOVA for the influence of saccade number on saccade amplitude for the first and the second saccade indicated significant effects ($ F(1,1790)=54.4 , p=2.5 \times 10^{-13}$), where the amplitude of the first saccade was larger than the amplitude of the second 
(Fig.~\ref{FigSaccAmp}b). 

In summary, forcing the observers to start exploration from an experimentally controlled initial position close to the border of the monitor resulted in a long first saccade. This was particularly true if the interesting image part was on the opposite side of the initial position. The longer initial saccade from right to left than vice versa is congruent to the left direction bias that has been found in various experiments \cite{dickinson2009spatial,foulsham2013leftward,ossandon2014spatial}. This result indicates that the leftward bias is not only present, if participants start observations from the center of the image.

\subsection{Saccade turning angle and its relation to amplitude}

\begin{figure}
\unitlength1mm
\begin{picture}(120,160)
\put(2,80){\includegraphics[width=16cm]{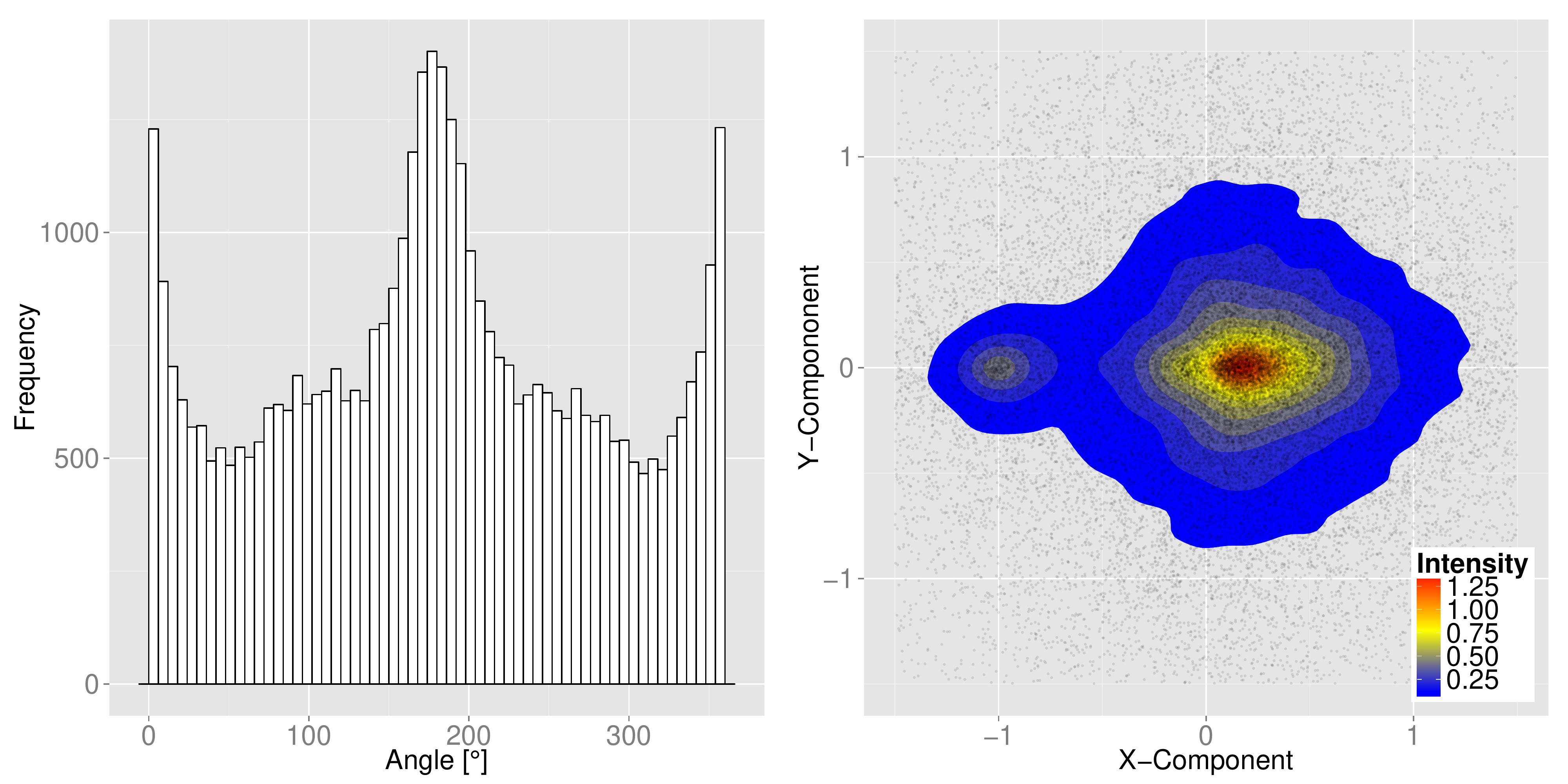}}
\put(0,88){(a)}
\put(81,88){(b)}
\put(42,0){\includegraphics[width=8cm]{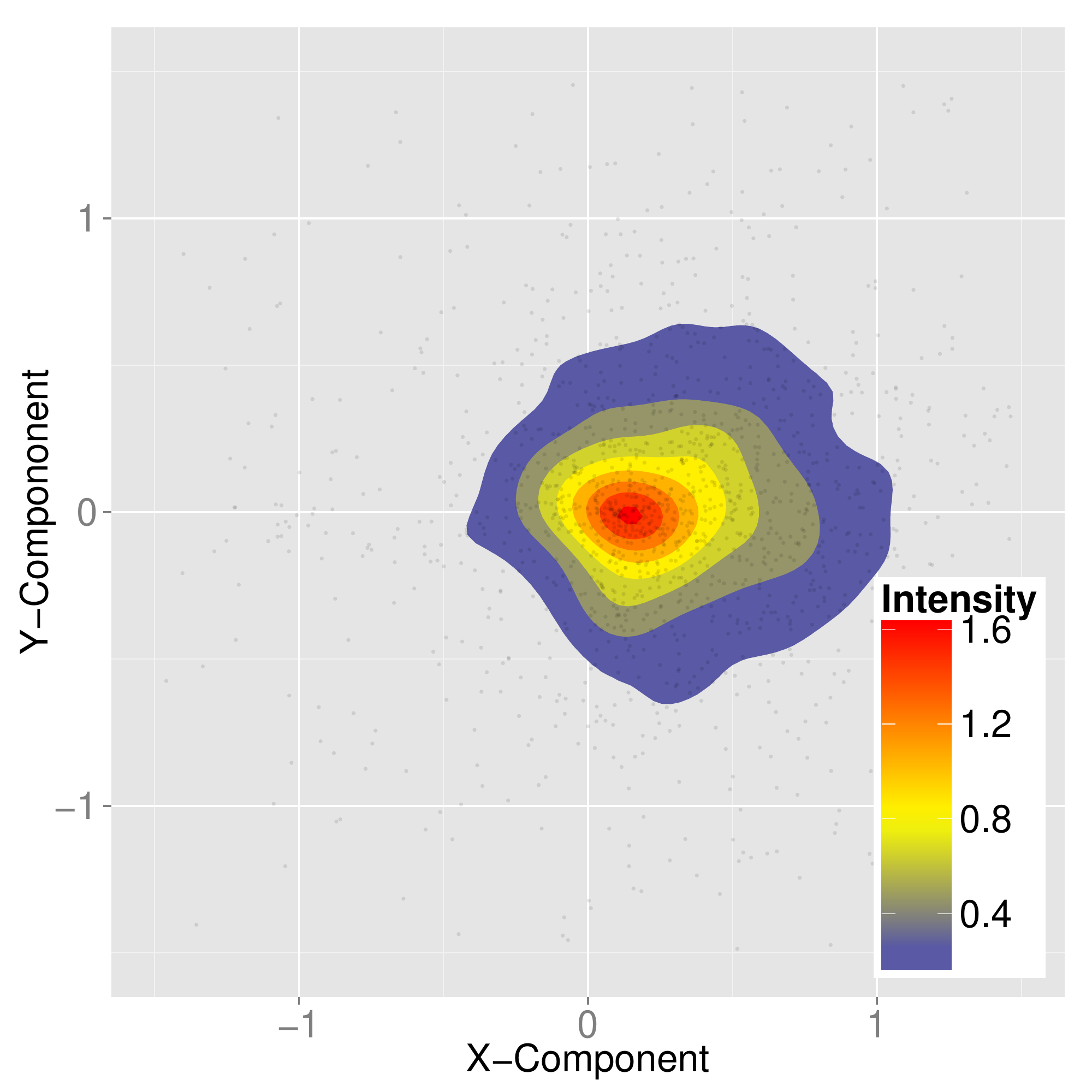}}
\put(40,8){(c)}

\end{picture}
\caption{\label{FigTurnAng}
Summary statistics for saccade turning angles. (a) The distribution of angles between two successive saccades is markedly peaked at the $0\degree$ and $180\degree$.
(b) Plot of the relation between saccade amplitude and turning angle contingent on parameters of the previous saccade. The previous saccadic endpoints are aligned to the origin. Saccade amplitudes were normalized to one and the saccade orientations were rotated to map the endpoints of a saccade with unit length to the point $(1,0)$. This representation shows that most saccades either travel in the same direction as the previous saccade, but with reduced saccade amplitude, or shift gaze back to the starting position of the previous saccade, i.e., the point $(-1;0)$. (c) same as (b) but only for the first two saccades. This shows that after the long first fixation return saccades back to this position are hardly present.}
\end{figure}
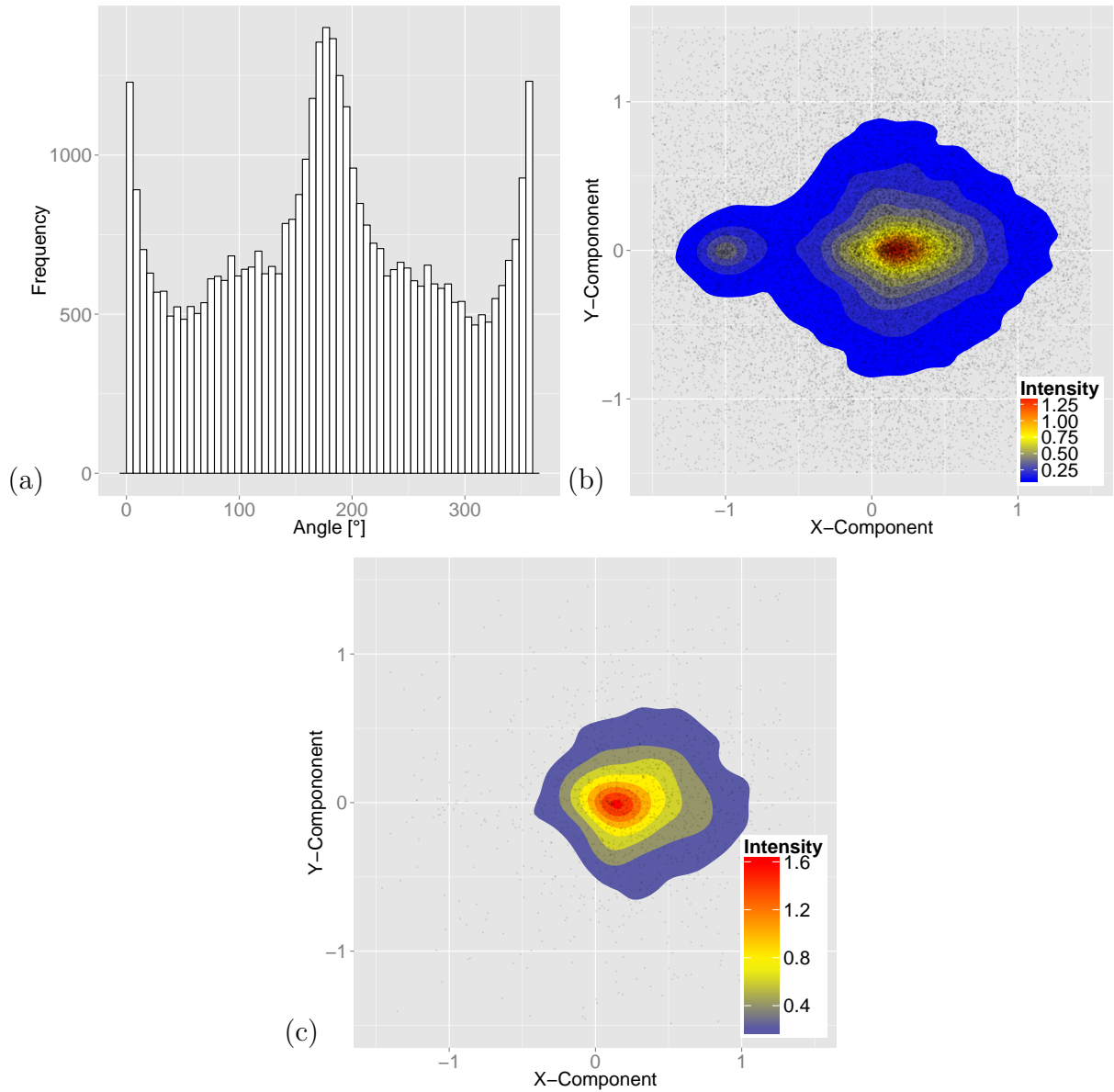

Statistically, most saccades are likely to follow the direction of previous saccades or shift gaze position back to the direction of the starting position of the previous saccade. The overall distribution of saccade turning angles between two subsequent saccades is characteristic for similar experiments in scene viewing (Fig.\ref{FigTurnAng}a) \cite{tatler2008systematic,smith2009facilitation}. 
Next, we constructed a conditional plot of saccade amplitude in relation to the previous saccade amplitude and orientation (Fig.\ref{FigTurnAng}b). The endpoint of the previous saccade was mapped to the origin of the coordinate system, saccade amplitude was normalized to the amplitude of the previous saccade, and the saccade orientation of the previous saccade was rotated to the right (or $180\degree$ orientation). In this representation, an endpoint at $(x;y)=(1;0)$ corresponds to a saccade that has the same length and orientation as the previous saccade (i.e., a turning angle of $180\degree$). The endpoint at $(x;y)=(-1;0)$ indicates that the saccade had the same amplitude as the previous saccade , but an opposite direction, which represents a perfect return saccade (i.e., a turning angle of $0\degree$). The high intensity at this point is consistent with earlier experiments that reported a large number of return saccades \cite{hooge2005inhibition,tatler2008systematic,smith2009facilitation}. 

Results from our analysis of turning angles and saccade amplitudes seem to be inconsistent with an inhibitory tagging mechanism. However, ruling out an inhibitory tagging mechanism based on these data would be premature, since inhibitory tagging could still be active, but not express in behavioral data on this level. Our analyses below will indicate a potential role of inhibitory tagging. Moreover, Figure \ref{FigTurnAng}c shows the same plot as figure \ref{FigTurnAng}b, but only for the first and second saccade. This plot indicates that a facilitation of return saccades did not appear during these first three fixations.

\subsection{Influence of starting position and image type on exploration behavior}
\begin{figure}
\unitlength1mm
\begin{picture}(150,115)
\put(0,0){\includegraphics[width=16cm]{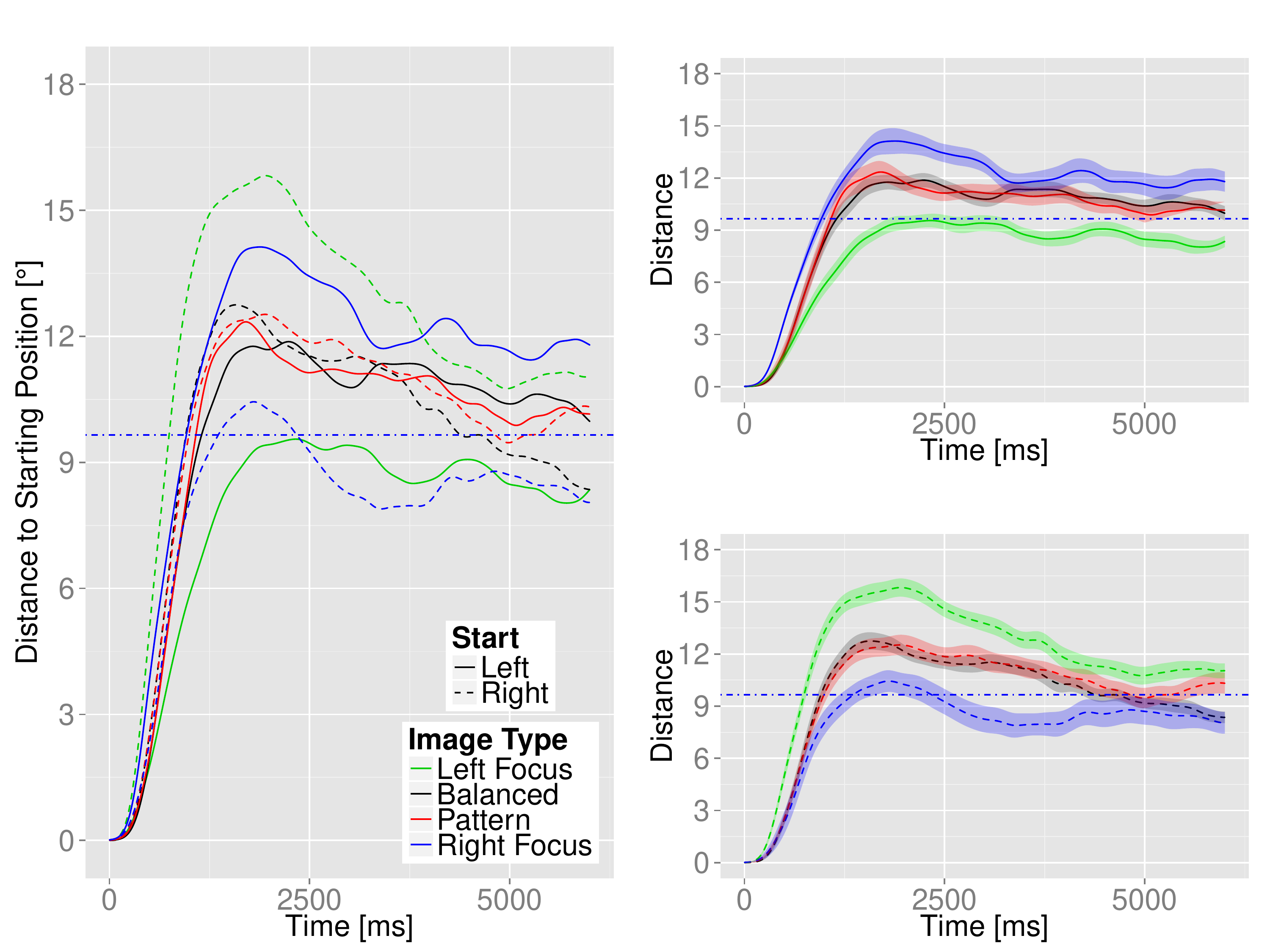}}
\put(0,1){(a)}
\put(80,1){(c)}
\put(80,59){(b)}
\end{picture}
\caption{\label{FigMHD}
Mean horizontal distance $X_{\rm MHD}(t)$ of gaze position at time $t$ from  starting position. (a) Almost all curves show an overshoot of the mean gaze position to the image side opposite to the starting position. (b) Curves from left starting positions with bootstrap-based confidence intervals. (c)  Curves from right starting positions with bootstrap-based confidence intervals.}
\end{figure}
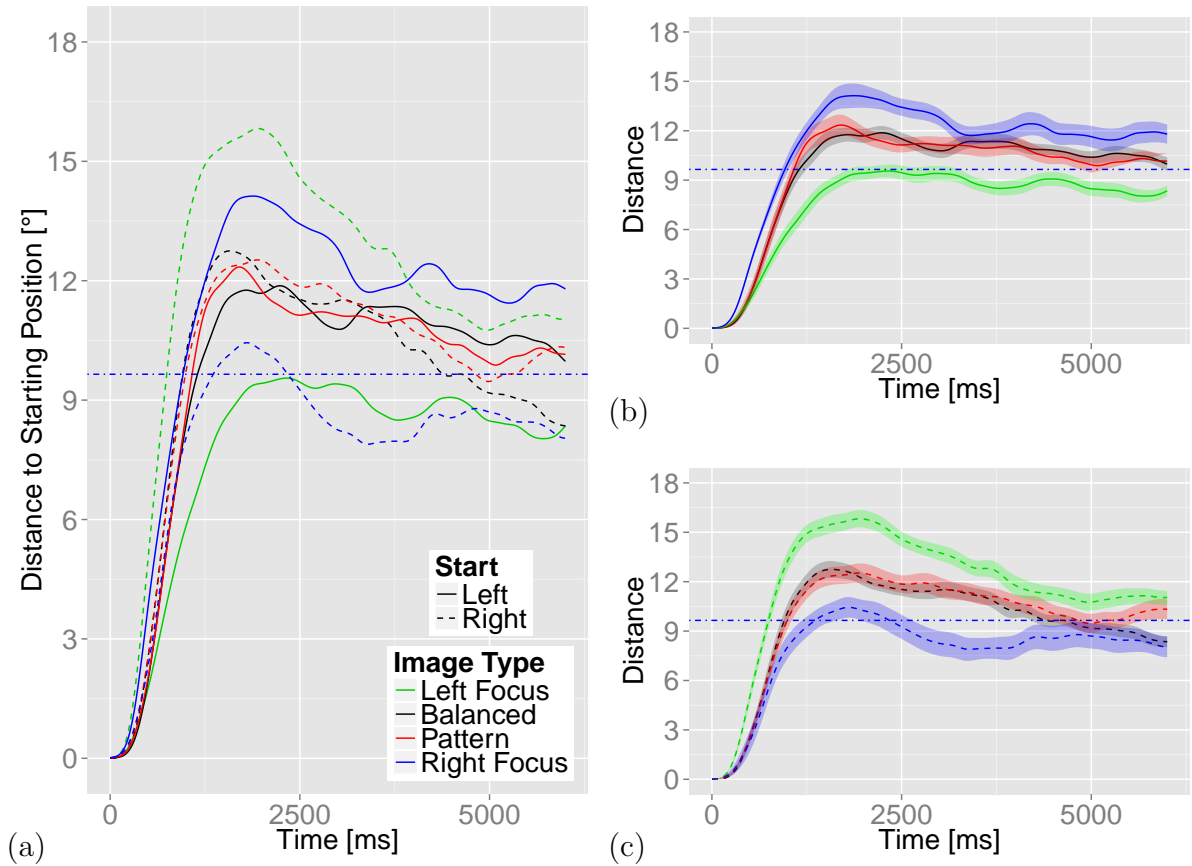

The most important aim of the current study was to investigate the influence of starting position on scanpath statistics. Therefore we introduced a measure of the mean horizontal distance (MHD) to the starting position at time $t$, denoted by $X_{\rm MHD}(t)$ (see {\sl Methods}). This measure was computed for each combination of image type and starting position (Fig.~\ref{FigMHD}a). The blue horizontal line indicates the horizontal center of the image. There are three important main effects in the plots of $X_{\rm MHD}(t)$. First, for the long term behavior in the balanced images and pattern images, $X_{\rm MHD}(t)$ approaches the midline, while there are obvious deviations for images with left or right focus. 

Second, the transient behavior induced by the starting position lasts to about 3~s to 5~s (depending on condition). This observation is in strong contrast to our finding that saccade amplitudes are only affected for the first saccade, which translates into a transient phase of the mean first fixation duration, equivalent to 609.01~ms. This untypically long first fixation indicates that it took the participants a long time to process that the fixation cross had disappeared. 

Third, in a time of approximately 1.5~s to 2~s almost all curves cross the midline and show a local maximum of MHD. The existence of such a maximum lends support for inhibition at the starting position, i.e., the eye is actively driven to the opposite image side. This is most evident for the conditions in which observers started in the image side opposite to the focus (starting from the right in left-focus images and starting from the left in right-focus images), but the effect is also visible for balanced images. Additionally, an interaction between image type and starting position is visible in Figure \ref{FigMHD}a. When observers started in the interesting side of an image, the final MHD to the starting position is smaller than for the balanced and pattern images and for the balanced and pattern images it is smaller than if participants started on the less salient image side. Graphs are cut off at $t=6000$~ms because after approximately 5 seconds the MHD reaches a stable value. 

Finally, we investigated the statistical reliability of our results via bootstrapping from $1\,000$ bootstrap samples of the 28 participants \cite{efron1994introduction}. The confidence intervals (Fig.~\ref{FigMHD}b,c) for the MHD curves $X_{\rm MHD}(t)$ were obtained by subtracting the subject mean and adding the overall mean to the samples as described by Cousineau \cite{cousineau2005confidence,loftus1994using} and taking the 2.5\% and 97.5\%  quantile of the MHD samples for the lower and upper bound. Confidence intervals show that MHD of left and right focus images differ for both starting positions significantly from the balanced images. Pattern images show almost the same MHD as balanced images.

\subsection{Comparison of experimental data with model simulations for scanpath statistics}
The analysis of the time-dependence of the mean horizontal distance to the starting position uncovered at least two unexpected results, (i) the observation of long transients and (ii) an overshoot component to the image side opposite to the starting position, even in the case of balanced images. To interpret the experimental findings we calculated the same statistics for computer-generated scanpaths from two statistical models, a saccadic momentum model and a dynamical model of scene exploration (see {\sl Methods}) and compared them to the experimental data. 

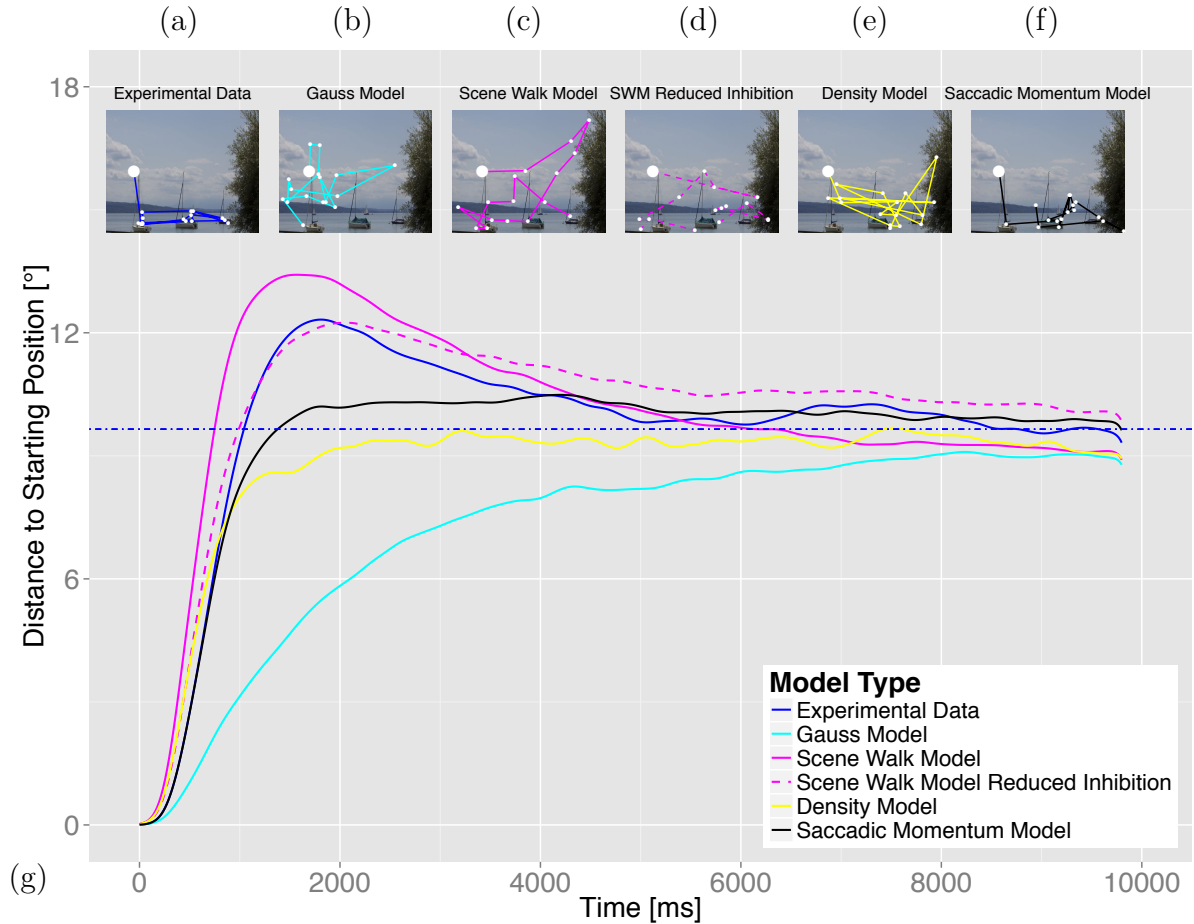
\begin{figure}
\unitlength1mm
\begin{picture}(150,115)
\put(0,-5){\includegraphics[width=16cm]{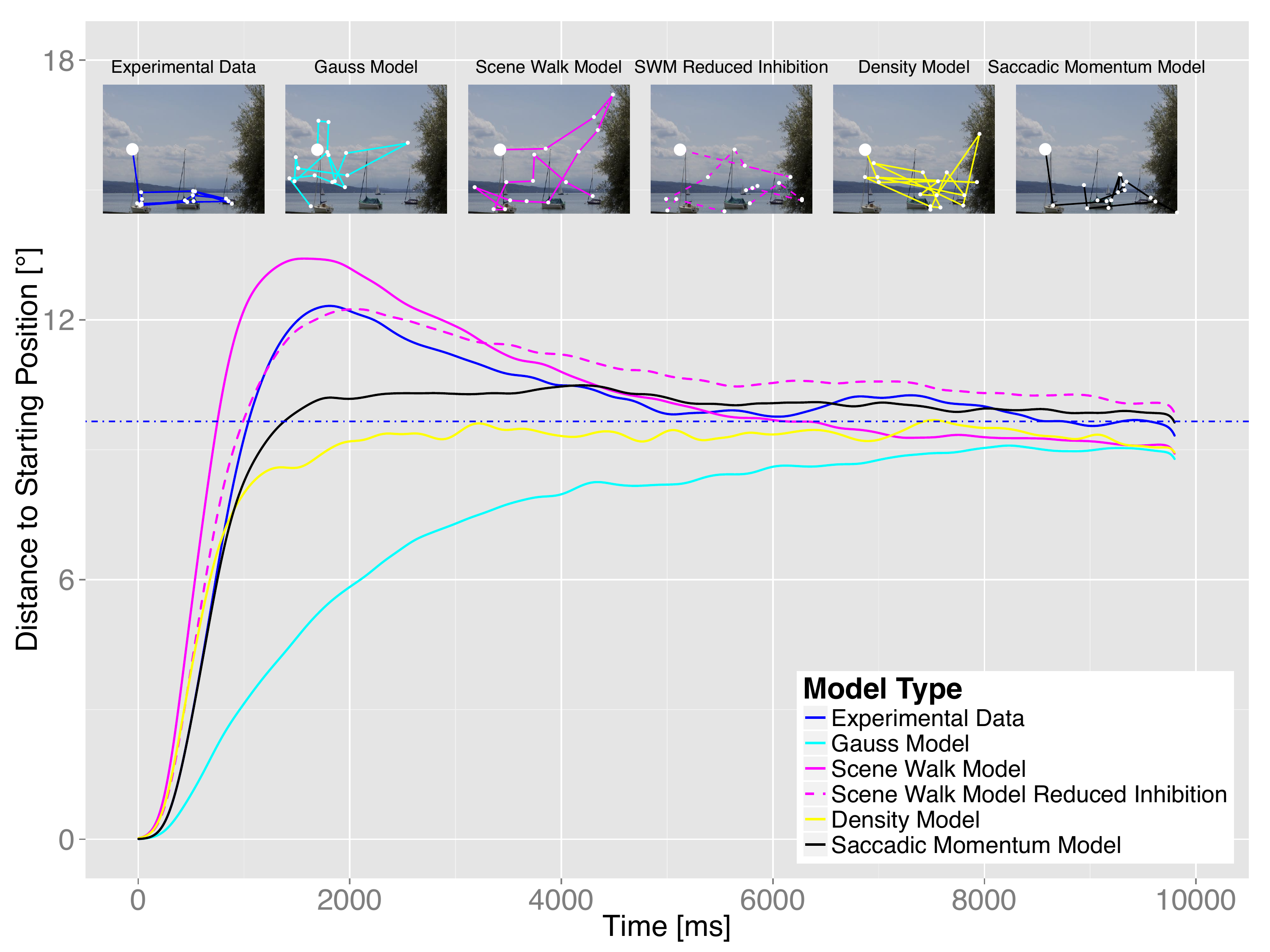}}
\put(20,115){(a)}
\put(43,115){(b)}
\put(66,115){(c)}
\put(89,115){(d)}
\put(112,115){(e)}
\put(135,115){(f)}
\put(0,1){(g)}
\end{picture}
\caption{\label{FigModels}
Comparison of mean horizontal distance of gaze from starting position for the experimental data and three models. Examples of scanpaths for (a) experimental data, (b) Gaussian weighted random sampling from density map, (c) SceneWalk model (Engbert et al., 2015) based on target selection from dynamic activation maps (d) SceneWalk model with a reduced strength of the inhibition map (e) random sampling from density map and (f) Saccadic momentum model. (g) Mean horizontal distance $X_{\rm MHD}(t)$ of gaze position at time $t$ shows that the qualitative behavior in the experimental data with an overshoot component to the image side opposite to the starting position is reproduced by the SceneWalk model that uses inhibitory tagging as a driving mechanism.}
\end{figure}

To illustrate model simulations we compare computer-generated scanpaths with experimental data. Figure \ref{FigModels}a-f shows one scanpath for the experimental data and each of the models. The big white marker on the right side indicates the initial fixation position. Simulated fixations from the gaussian model (bright blue path) remain close to the current fixation position due to a limited attentional span. The density map model (yellow path) produces longer saccade amplitudes than observed in the experiment and cannot reproduce the typical distribution of saccade amplitudes. The SceneWalk model (pink paths) produces scanpaths that are statistically more similar to the experimental data than the two random-sampling models. The saccadic momentum model shows similar scanpath statistics as the data, as angles and amplitudes are sampled from the data and the first two fixations are equal. Next, we analysed the mean horizontal distance for all simulations. 

Figure \ref{FigModels}g presents the mean horizontal distance for experimental data in comparison to data generated by the computational models, averaged over image types and starting positions. The most straightforward model is based on random sampling from the experimentally observed fixation density. The MHD curve indicates that this model cannot produce an overshoot to the image side opposite to the starting position. A model that samples fixation positions randomly from a 2D gaussian-weighted density map is too slow in leaving the starting position, which is indicated by the shallow slope of the MHD curve. While the Gaussian-weighted model is psychologically more plausible because of its limited attentional span, the density map model is more compatible with the experimental data. Although the saccadic momentum model contains the same initial three fixations as the data, it can not reproduce the overshoot from the data. In contrast to the other models, the dynamical SceneWalk model \cite{engbert2015spatial} can reproduce the overshoot component of the MHD curves in the time interval between 1.5~s and 2~s. The SceneWalk model uses inhibitory tagging that drives the eye towards the image side opposite of the starting position by inhibiting selection of saccade targets close to the starting position of the scanpath. Since model parameters of the SceneWalk model were taken from the published version and not adjusted to the current experimental data, we changed the exponent of the inhibition map from $\gamma=.3$ to $.2$ (see Eq. \ref{Eq_Pot} in Methods) in a second simulation. This parameter was also adjusted by hand in the previous version of the model. We see that the overshoot of the MHD curve from the Scene Walk Model with an adjusted inhibition size is in good agreement with the overshoot observed in the experimental data. These simulations suggest that the overshoot produced by the model is primarily caused by the inhibitory tagging mechanism. 
 
\begin{figure}
\unitlength1mm
\begin{picture}(150,115)
\put(2,-2){\includegraphics[width=16cm]{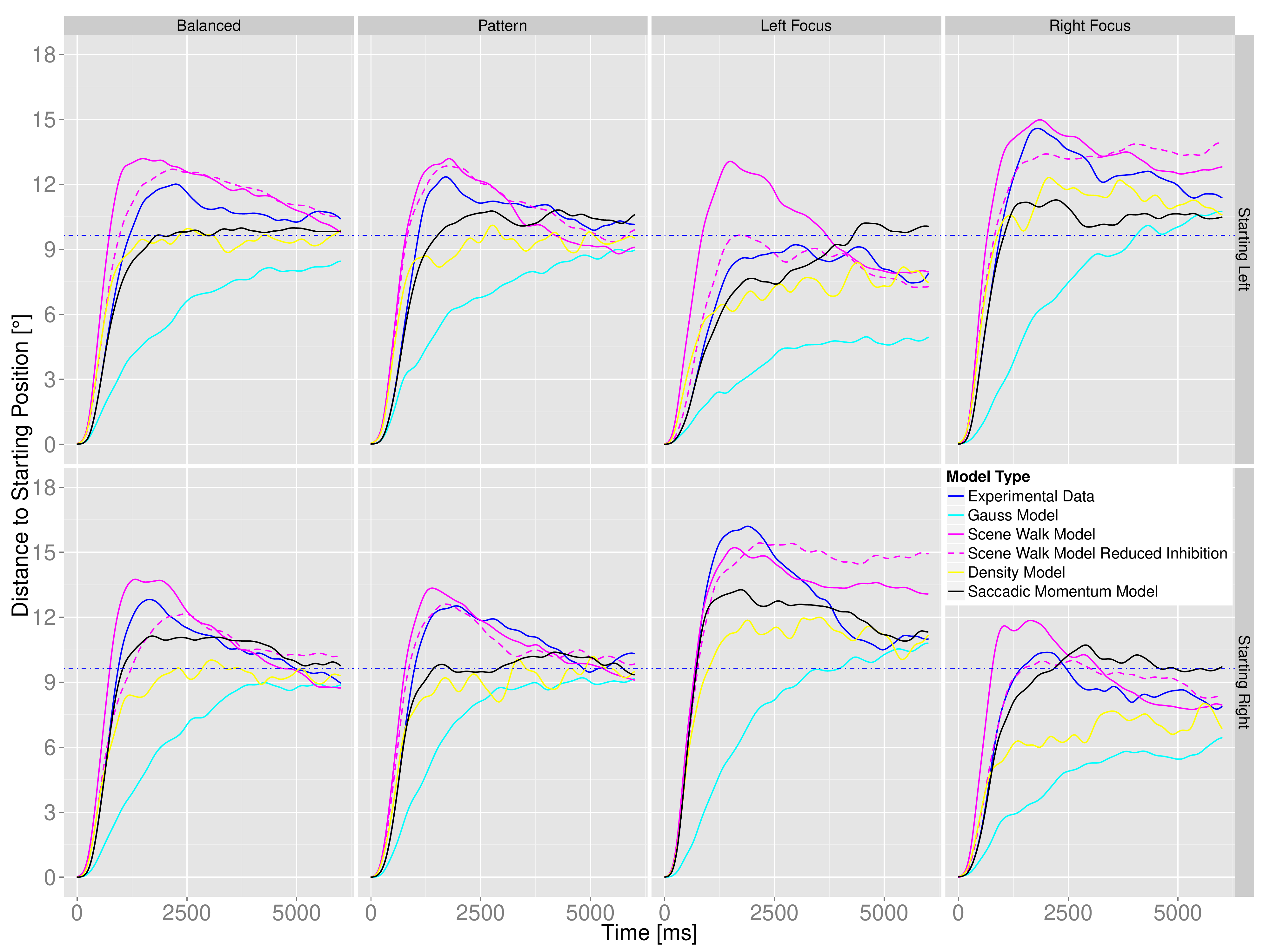}}
\end{picture}
\caption{\label{FigModelsDetail}
Mean horizontal distance to the starting position for all 8 combinations of image type and starting position, for the 4 different scanpath models and experimental data. In all but one condition (left-focus image with left starting position), an overshoot of the mean position to the image side opposite to the starting position is visible in the experimental data. This overshoot was reproduced by the dynamical SceneWalk model that implements inhibitory tagging.}
\end{figure}

While the analysis in Figure \ref{FigModels} is based on averaging over experimental conditions, Figure \ref{FigModelsDetail} shows the MHD for each of the different conditions for experimental data and model simulations. Across all conditions, the Gaussian-weighted density map is too slow in approaching the image side opposite to the starting position, while random sampling from the density map gives an acceptable fit to some experimental conditions. Initially the saccadic momentum model equals the experimental data curve, because they are the same for the first three fixations. Afterwards it moves too slowly away from the starting position and does not show an overshoot.

\begin{figure}
\unitlength1mm
\begin{picture}(150,115)
\put(2,-2){\includegraphics[width=16cm]{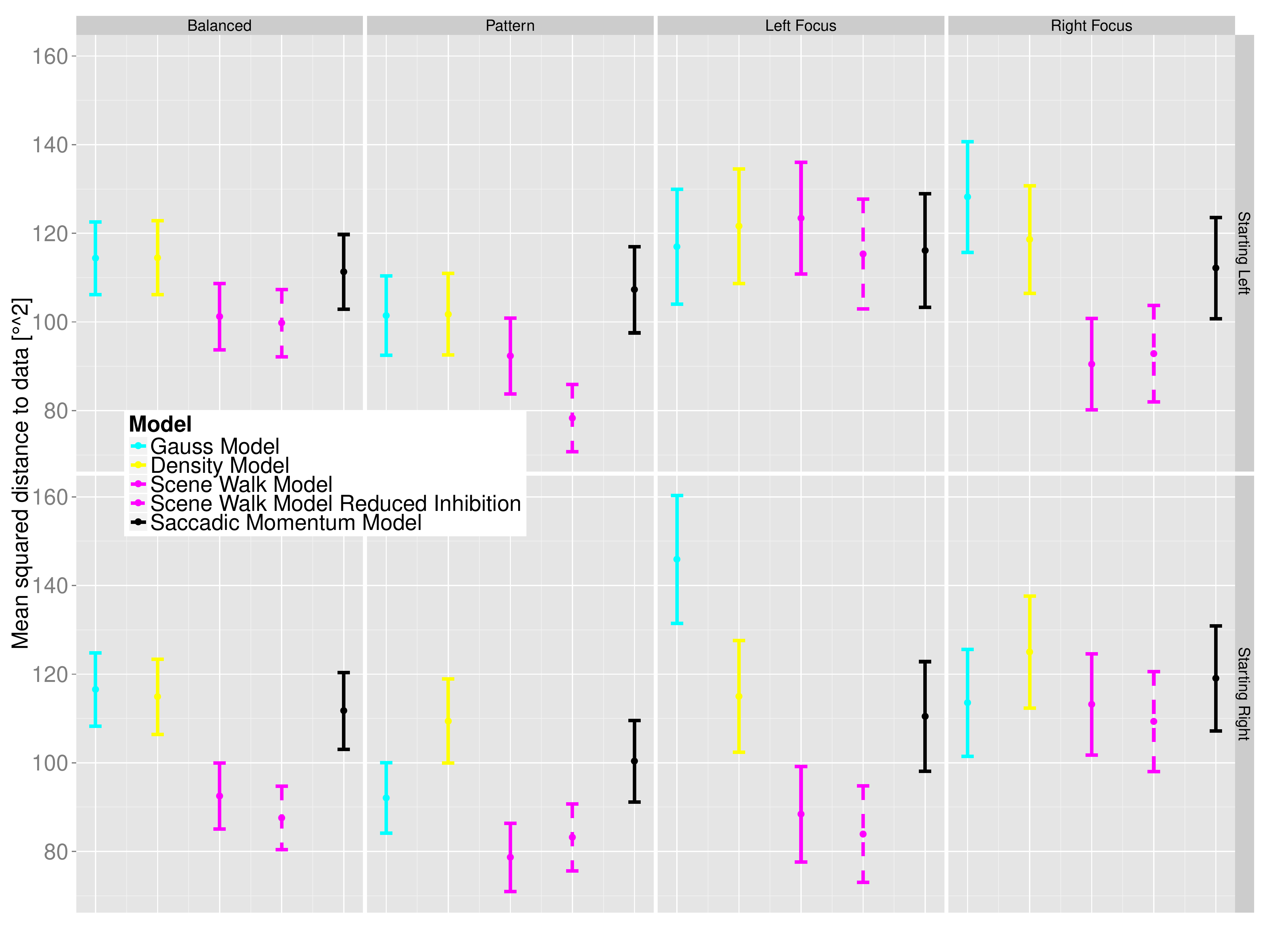}}
\end{picture}
\caption{\label{FigModelsDetail2}
Mean squared distance of the model curves to the experimental data between 1.36 and 5 seconds. Either the original or the adjusted version of the Scene walk model perform best in all conditions. Errorbars represent the standard error of the mean.}
\end{figure}

While overshoots in the SceneWalk model are often too strong compared to experimental data, inhibitory tagging as implemented in our model is a viable computational mechanism to explain the existence of an overshoot of eye position to image side opposite to the starting position. Quantitave evaluation is shown in Fig.\ref{FigModelsDetail2}. The saccadic momentum model keeps the first 3 fixations from the experimental data, which relates to a mean fixation duration of 1360~ms. Therefore, we investigated the time window from 1.36~s to 5~s (no systematic effects can be seen after 5~s). Across all models, the orignial SceneWalk model or the SceneWalk model with the adjusted inhibition strength is closest to the experimental data. Thus, there is a clear trend that it outperforms the saccadic momentum model and the statistical models with respect to the mean horizontal distance.

\subsection{Discussion}

In an eye tracking experiment we investigated the influence of experimentally manipulated starting positions on scanpath behavior in human observers. Most important effects were observed in the temporal evolution of the mean horizontal distance (MHD) to the starting position. First, we found unexpectedly long transients in mean eye position. It took up to 5 seconds for gaze of human observers to reach the final average fixation position. This is a lot longer than the saccade amplitude effects, which were limited to the very first saccade of the observers' scanpaths. Second, for almost all experimental conditions the MHD over time is characterized by a strong overshoot of the midline into the image side opposite to the starting position before reaching a stable value. This effect lends support to a foraging strategy that actively moves the gaze to unexplored image regions although on a shorter time scale a high number of return saccades suggests the opposite.

Next, we analyzed computational models that can predict human scanpaths. Random sampling from the empirical saliency map (i.e., assuming a `perfect' saliency model) does not replicate human behavior, since the overshoot to the opposite side of the image cannot be reproduced. Moreover, such a model is psychologically highly implausible because of the missing effect of degraded visual acuity towards the periphery of the visual field. However, an augmented model, i.e., a combination of the density map with a Gaussian attention window representing the fall-off of acuity to the periphery, performs even worse compared to random sampling from the empirical saliency map. We conclude from these results that an active mechanism driving the eyes away from the starting position is necessary to explain scanpath statistics of the time-dependence of mean horizontal distance.

Given the above experimental results, we were looking for potential principles of eye guidance that drive the trajectory faster away from the current fixation position than a simple random process. We investigated two principals in computational models: saccadic momentum \cite{smith2009facilitation,wilming2013saccadic} and spatial inhibitory tagging \cite{itti1998model,le2015saccadic}.

A model based on saccadic momentum that samples from the joint probability distribution of saccade angles and amplitudes and keeps the first two saccades observed in the experiments. As a trivial result, the first three fixations in each trial from the simulations fit the experimental data better than any other model, however, the model did not reproduce the overshoot component to the opposite image side. We also used the SceneWalk model (Engbert et al., 2015), a dynamical model for eye-movement control in scenes that aimed at reproducing first- and second-order statistics, i.e., densities of fixations and clustering, respectively. The SceneWalk model uses inhibitory tagging, a mechanism motivated by the findings on inhibition of return \cite{posner1984components,posner1985inhibition,klein1988inhibitory}. We demonstrated that the SceneWalk model generates the overshoot effect for MHD via inhibitory tagging in contrast to the two random-sampling models or the saccadic momentum model. With the parameters that were fitted from a different experiment, the SceneWalk model produced MHD curves that were qualitatively similar to the curves computed from the experimental data. The fit to the data was improved by reducing the influence of the inhibition map by reduction of a parameter controlling the strength of inhibition. A facilitation of return \cite{smith2009facilitation} back to the starting position was not observed.

Inspections of a left-focus image from a starting position on the left show different dynamics of the mean horizontal distance compared to all other conditions. This could be due to a stronger directional bias in left-focus images than in right-focus images in our experimental material. It is also possible that there is a general tendency to first look at the left image side and then scan to the right---a tendency that has been found earlier in scene viewing \cite{dickinson2009spatial,ossandon2014spatial}, pattern exploration \cite{abed1991cultural} and face viewing \cite{guo2009left}---which is congruent to the reading direction of our participants. A dynamical model of eye guidance might perform better with an additional Bayesian-type prior probability implementing a leftward bias and a center bias for initial saccades. Thus, our results emphasize the need for more advanced dynamical models of scanpath generation. 

\section{Conclusion}
The experimental manipulation of starting position exerts a strong and long lasting influence on scanpaths during scene exploration. Using computational models, we demonstrate that a model with inhibitory tagging can explain the mean overshoot of gaze position to the image side opposite to the starting position whilst simple statistical models as well as a saccadic momentum model can not reproduce this overshoot. These results lend support to inhibitory tagging as a dynamical principle of saccade planning during scene viewing.

\section{Acknowledgements}
This work was supported by Deutsche Forschungsgemeinschaft (grants EN 471/13–1 and WI 2103/4–1 to R. E. and F. A. W., resp.).

\bibliographystyle{apalike}
\bibliography{Library}

\begin{thebibliography}{}

\bibitem [\protect \citeauthoryear {%
Abed%
}{%
Abed%
}{%
{\protect \APACyear {1991}}%
}]{%
abed1991cultural}
\APACinsertmetastar {%
abed1991cultural}%
\begin{APACrefauthors}%
Abed, F.%
\end{APACrefauthors}%
\unskip\
\newblock
\APACrefYearMonthDay{1991}{}{}.
\newblock
{\BBOQ}\APACrefatitle {Cultural influences on visual scanning patterns}
  {Cultural influences on visual scanning patterns}.{\BBCQ}
\newblock
\APACjournalVolNumPages{Journal of Cross-Cultural Psychology}{22}{4}{525--534}.
\PrintBackRefs{\CurrentBib}

\bibitem [\protect \citeauthoryear {%
Baddeley%
\ \BBA {} Turner%
}{%
Baddeley%
\ \BBA {} Turner%
}{%
{\protect \APACyear {2005}}%
}]{%
SpatStat}
\APACinsertmetastar {%
SpatStat}%
\begin{APACrefauthors}%
Baddeley, A.%
\BCBT {}\ \BBA {} Turner, R.%
\end{APACrefauthors}%
\unskip\
\newblock
\APACrefYearMonthDay{2005}{}{}.
\newblock
{\BBOQ}\APACrefatitle {{spatstat}: An {R} Package for Analyzing Spatial Point
  Patterns} {{spatstat}: An {R} package for analyzing spatial point
  patterns}.{\BBCQ}
\newblock
\APACjournalVolNumPages{Journal of Statistical Software}{12}{6}{1--42}.
\newblock
\begin{APACrefURL} \url{http://www.jstatsoft.org/v12/i06/} \end{APACrefURL}
\PrintBackRefs{\CurrentBib}

\bibitem [\protect \citeauthoryear {%
Bates%
, Maechler%
, Bolker%
\BCBL {}\ \BBA {} Walker%
}{%
Bates%
\ \protect \BOthers {.}}{%
{\protect \APACyear {2013}}%
}]{%
bates2013lme4}
\APACinsertmetastar {%
bates2013lme4}%
\begin{APACrefauthors}%
Bates, D.%
, Maechler, M.%
, Bolker, B.%
\BCBL {}\ \BBA {} Walker, S.%
\end{APACrefauthors}%
\unskip\
\newblock
\APACrefYearMonthDay{2013}{}{}.
\newblock
{\BBOQ}\APACrefatitle {lme4: Linear mixed-effects models using Eigen and S4}
  {lme4: Linear mixed-effects models using eigen and s4}.{\BBCQ}
\newblock
\APACjournalVolNumPages{R package version}{1}{4}{}.
\PrintBackRefs{\CurrentBib}

\bibitem [\protect \citeauthoryear {%
Bays%
\ \BBA {} Husain%
}{%
Bays%
\ \BBA {} Husain%
}{%
{\protect \APACyear {2012}}%
}]{%
bays2012active}
\APACinsertmetastar {%
bays2012active}%
\begin{APACrefauthors}%
Bays, P\BPBI M.%
\BCBT {}\ \BBA {} Husain, M.%
\end{APACrefauthors}%
\unskip\
\newblock
\APACrefYearMonthDay{2012}{}{}.
\newblock
{\BBOQ}\APACrefatitle {Active inhibition and memory promote exploration and
  search of natural scenes} {Active inhibition and memory promote exploration
  and search of natural scenes}.{\BBCQ}
\newblock
\APACjournalVolNumPages{Journal of vision}{12}{8}{8}.
\PrintBackRefs{\CurrentBib}

\bibitem [\protect \citeauthoryear {%
Bindemann%
}{%
Bindemann%
}{%
{\protect \APACyear {2010}}%
}]{%
bindemann2010scene}
\APACinsertmetastar {%
bindemann2010scene}%
\begin{APACrefauthors}%
Bindemann, M.%
\end{APACrefauthors}%
\unskip\
\newblock
\APACrefYearMonthDay{2010}{}{}.
\newblock
{\BBOQ}\APACrefatitle {Scene and screen center bias early eye movements in
  scene viewing} {Scene and screen center bias early eye movements in scene
  viewing}.{\BBCQ}
\newblock
\APACjournalVolNumPages{Vision research}{50}{23}{2577--2587}.
\PrintBackRefs{\CurrentBib}

\bibitem [\protect \citeauthoryear {%
Borji%
\ \BBA {} Itti%
}{%
Borji%
\ \BBA {} Itti%
}{%
{\protect \APACyear {2013}}%
}]{%
borji2013state}
\APACinsertmetastar {%
borji2013state}%
\begin{APACrefauthors}%
Borji, A.%
\BCBT {}\ \BBA {} Itti, L.%
\end{APACrefauthors}%
\unskip\
\newblock
\APACrefYearMonthDay{2013}{}{}.
\newblock
{\BBOQ}\APACrefatitle {State-of-the-art in visual attention modeling}
  {State-of-the-art in visual attention modeling}.{\BBCQ}
\newblock
\APACjournalVolNumPages{Pattern Analysis and Machine Intelligence, IEEE
  Transactions on}{35}{1}{185--207}.
\PrintBackRefs{\CurrentBib}

\bibitem [\protect \citeauthoryear {%
Bylinskii%
\ \protect \BOthers {.}}{%
Bylinskii%
\ \protect \BOthers {.}}{%
{\protect \APACyear {2015}}%
}]{%
mit-saliency-benchmark}
\APACinsertmetastar {%
mit-saliency-benchmark}%
\begin{APACrefauthors}%
Bylinskii, Z.%
, Judd, T.%
, Borji, A.%
, Itti, L.%
, Durand, F.%
, Oliva, A.%
\BCBL {}\ \BBA {} Torralba, A.%
\end{APACrefauthors}%
\unskip\
\newblock
\APACrefYearMonthDay{2015}{}{}.
\newblock
\APACrefbtitle {MIT Saliency Benchmark.} {Mit saliency benchmark.}
\newblock
\APAChowpublished {http://saliency.mit.edu/}.
\PrintBackRefs{\CurrentBib}

\bibitem [\protect \citeauthoryear {%
Castelhano%
, Mack%
\BCBL {}\ \BBA {} Henderson%
}{%
Castelhano%
\ \protect \BOthers {.}}{%
{\protect \APACyear {2009}}%
}]{%
castelhano2009viewing}
\APACinsertmetastar {%
castelhano2009viewing}%
\begin{APACrefauthors}%
Castelhano, M\BPBI S.%
, Mack, M\BPBI L.%
\BCBL {}\ \BBA {} Henderson, J\BPBI M.%
\end{APACrefauthors}%
\unskip\
\newblock
\APACrefYearMonthDay{2009}{}{}.
\newblock
{\BBOQ}\APACrefatitle {Viewing task influences eye movement control during
  active scene perception} {Viewing task influences eye movement control during
  active scene perception}.{\BBCQ}
\newblock
\APACjournalVolNumPages{Journal of Vision}{9}{3}{6}.
\PrintBackRefs{\CurrentBib}

\bibitem [\protect \citeauthoryear {%
Cerf%
, Harel%
, Einh{\"a}user%
\BCBL {}\ \BBA {} Koch%
}{%
Cerf%
\ \protect \BOthers {.}}{%
{\protect \APACyear {2008}}%
}]{%
cerf2008predicting}
\APACinsertmetastar {%
cerf2008predicting}%
\begin{APACrefauthors}%
Cerf, M.%
, Harel, J.%
, Einh{\"a}user, W.%
\BCBL {}\ \BBA {} Koch, C.%
\end{APACrefauthors}%
\unskip\
\newblock
\APACrefYearMonthDay{2008}{}{}.
\newblock
{\BBOQ}\APACrefatitle {Predicting human gaze using low-level saliency combined
  with face detection} {Predicting human gaze using low-level saliency combined
  with face detection}.{\BBCQ}
\newblock
\BIn{} \APACrefbtitle {Advances in neural information processing systems}
  {Advances in neural information processing systems}\ (\BPGS\ 241--248).
\PrintBackRefs{\CurrentBib}

\bibitem [\protect \citeauthoryear {%
Cousineau%
}{%
Cousineau%
}{%
{\protect \APACyear {2005}}%
}]{%
cousineau2005confidence}
\APACinsertmetastar {%
cousineau2005confidence}%
\begin{APACrefauthors}%
Cousineau, D.%
\end{APACrefauthors}%
\unskip\
\newblock
\APACrefYearMonthDay{2005}{}{}.
\newblock
{\BBOQ}\APACrefatitle {Confidence intervals in within-subject designs: A
  simpler solution to Loftus and Masson's method} {Confidence intervals in
  within-subject designs: A simpler solution to loftus and masson's
  method}.{\BBCQ}
\newblock
\APACjournalVolNumPages{Tutorials in Quantitative Methods for
  Psychology}{1}{1}{42--45}.
\PrintBackRefs{\CurrentBib}

\bibitem [\protect \citeauthoryear {%
Deubel%
\ \BBA {} Schneider%
}{%
Deubel%
\ \BBA {} Schneider%
}{%
{\protect \APACyear {1996}}%
}]{%
deubel1996saccade}
\APACinsertmetastar {%
deubel1996saccade}%
\begin{APACrefauthors}%
Deubel, H.%
\BCBT {}\ \BBA {} Schneider, W\BPBI X.%
\end{APACrefauthors}%
\unskip\
\newblock
\APACrefYearMonthDay{1996}{}{}.
\newblock
{\BBOQ}\APACrefatitle {Saccade target selection and object recognition:
  Evidence for a common attentional mechanism} {Saccade target selection and
  object recognition: Evidence for a common attentional mechanism}.{\BBCQ}
\newblock
\APACjournalVolNumPages{Vision research}{36}{12}{1827--1837}.
\PrintBackRefs{\CurrentBib}

\bibitem [\protect \citeauthoryear {%
Dickinson%
\ \BBA {} Intraub%
}{%
Dickinson%
\ \BBA {} Intraub%
}{%
{\protect \APACyear {2009}}%
}]{%
dickinson2009spatial}
\APACinsertmetastar {%
dickinson2009spatial}%
\begin{APACrefauthors}%
Dickinson, C\BPBI A.%
\BCBT {}\ \BBA {} Intraub, H.%
\end{APACrefauthors}%
\unskip\
\newblock
\APACrefYearMonthDay{2009}{}{}.
\newblock
{\BBOQ}\APACrefatitle {Spatial asymmetries in viewing and remembering scenes:
  Consequences of an attentional bias?} {Spatial asymmetries in viewing and
  remembering scenes: Consequences of an attentional bias?}{\BBCQ}
\newblock
\APACjournalVolNumPages{Attention, Perception, \&
  Psychophysics}{71}{6}{1251--1262}.
\PrintBackRefs{\CurrentBib}

\bibitem [\protect \citeauthoryear {%
Efron%
\ \BBA {} Tibshirani%
}{%
Efron%
\ \BBA {} Tibshirani%
}{%
{\protect \APACyear {1994}}%
}]{%
efron1994introduction}
\APACinsertmetastar {%
efron1994introduction}%
\begin{APACrefauthors}%
Efron, B.%
\BCBT {}\ \BBA {} Tibshirani, R\BPBI J.%
\end{APACrefauthors}%
\unskip\
\newblock
\APACrefYear{1994}.
\newblock
\APACrefbtitle {An introduction to the bootstrap} {An introduction to the
  bootstrap}.
\newblock
\APACaddressPublisher{}{CRC press}.
\PrintBackRefs{\CurrentBib}

\bibitem [\protect \citeauthoryear {%
Engbert%
\ \BBA {} Kliegl%
}{%
Engbert%
\ \BBA {} Kliegl%
}{%
{\protect \APACyear {2003}}%
}]{%
engbert2003microsaccades}
\APACinsertmetastar {%
engbert2003microsaccades}%
\begin{APACrefauthors}%
Engbert, R.%
\BCBT {}\ \BBA {} Kliegl, R.%
\end{APACrefauthors}%
\unskip\
\newblock
\APACrefYearMonthDay{2003}{}{}.
\newblock
{\BBOQ}\APACrefatitle {Microsaccades uncover the orientation of covert
  attention} {Microsaccades uncover the orientation of covert
  attention}.{\BBCQ}
\newblock
\APACjournalVolNumPages{Vision research}{43}{9}{1035--1045}.
\PrintBackRefs{\CurrentBib}

\bibitem [\protect \citeauthoryear {%
Engbert%
\ \BBA {} Mergenthaler%
}{%
Engbert%
\ \BBA {} Mergenthaler%
}{%
{\protect \APACyear {2006}}%
}]{%
engbert2006microsaccades}
\APACinsertmetastar {%
engbert2006microsaccades}%
\begin{APACrefauthors}%
Engbert, R.%
\BCBT {}\ \BBA {} Mergenthaler, K.%
\end{APACrefauthors}%
\unskip\
\newblock
\APACrefYearMonthDay{2006}{}{}.
\newblock
{\BBOQ}\APACrefatitle {Microsaccades are triggered by low retinal image slip}
  {Microsaccades are triggered by low retinal image slip}.{\BBCQ}
\newblock
\APACjournalVolNumPages{Proceedings of the National Academy of
  Sciences}{103}{18}{7192--7197}.
\PrintBackRefs{\CurrentBib}

\bibitem [\protect \citeauthoryear {%
Engbert%
, Trukenbrod%
, Barthelm{\'e}%
\BCBL {}\ \BBA {} Wichmann%
}{%
Engbert%
\ \protect \BOthers {.}}{%
{\protect \APACyear {2015}}%
}]{%
engbert2015spatial}
\APACinsertmetastar {%
engbert2015spatial}%
\begin{APACrefauthors}%
Engbert, R.%
, Trukenbrod, H\BPBI A.%
, Barthelm{\'e}, S.%
\BCBL {}\ \BBA {} Wichmann, F\BPBI A.%
\end{APACrefauthors}%
\unskip\
\newblock
\APACrefYearMonthDay{2015}{}{}.
\newblock
{\BBOQ}\APACrefatitle {Spatial statistics and attentional dynamics in scene
  viewing} {Spatial statistics and attentional dynamics in scene
  viewing}.{\BBCQ}
\newblock
\APACjournalVolNumPages{Journal of vision}{15}{1}{14}.
\PrintBackRefs{\CurrentBib}

\bibitem [\protect \citeauthoryear {%
Findlay%
\ \BBA {} Gilchrist%
}{%
Findlay%
\ \BBA {} Gilchrist%
}{%
{\protect \APACyear {2003}}%
}]{%
findlayactive}
\APACinsertmetastar {%
findlayactive}%
\begin{APACrefauthors}%
Findlay, J\BPBI M.%
\BCBT {}\ \BBA {} Gilchrist, I\BPBI D.%
\end{APACrefauthors}%
\unskip\
\newblock
\APACrefYear{2003}.
\newblock
\APACrefbtitle {Active vision: The psychology of looking and seeing} {Active
  vision: The psychology of looking and seeing}.
\newblock
\APACaddressPublisher{}{Oxford: Oxford University Press}.
\PrintBackRefs{\CurrentBib}

\bibitem [\protect \citeauthoryear {%
Foulsham%
, Gray%
, Nasiopoulos%
\BCBL {}\ \BBA {} Kingstone%
}{%
Foulsham%
\ \protect \BOthers {.}}{%
{\protect \APACyear {2013}}%
}]{%
foulsham2013leftward}
\APACinsertmetastar {%
foulsham2013leftward}%
\begin{APACrefauthors}%
Foulsham, T.%
, Gray, A.%
, Nasiopoulos, E.%
\BCBL {}\ \BBA {} Kingstone, A.%
\end{APACrefauthors}%
\unskip\
\newblock
\APACrefYearMonthDay{2013}{}{}.
\newblock
{\BBOQ}\APACrefatitle {Leftward biases in picture scanning and line bisection:
  A gaze-contingent window study} {Leftward biases in picture scanning and line
  bisection: A gaze-contingent window study}.{\BBCQ}
\newblock
\APACjournalVolNumPages{Vision research}{78}{}{14--25}.
\PrintBackRefs{\CurrentBib}

\bibitem [\protect \citeauthoryear {%
Foulsham%
\ \BBA {} Kingstone%
}{%
Foulsham%
\ \BBA {} Kingstone%
}{%
{\protect \APACyear {2010}}%
}]{%
foulsham2010asymmetries}
\APACinsertmetastar {%
foulsham2010asymmetries}%
\begin{APACrefauthors}%
Foulsham, T.%
\BCBT {}\ \BBA {} Kingstone, A.%
\end{APACrefauthors}%
\unskip\
\newblock
\APACrefYearMonthDay{2010}{}{}.
\newblock
{\BBOQ}\APACrefatitle {Asymmetries in the direction of saccades during
  perception of scenes and fractals: Effects of image type and image features}
  {Asymmetries in the direction of saccades during perception of scenes and
  fractals: Effects of image type and image features}.{\BBCQ}
\newblock
\APACjournalVolNumPages{Vision Research}{50}{8}{779--795}.
\PrintBackRefs{\CurrentBib}

\bibitem [\protect \citeauthoryear {%
Gilchrist%
\ \BBA {} Harvey%
}{%
Gilchrist%
\ \BBA {} Harvey%
}{%
{\protect \APACyear {2000}}%
}]{%
gilchrist2000refixation}
\APACinsertmetastar {%
gilchrist2000refixation}%
\begin{APACrefauthors}%
Gilchrist, I\BPBI D.%
\BCBT {}\ \BBA {} Harvey, M.%
\end{APACrefauthors}%
\unskip\
\newblock
\APACrefYearMonthDay{2000}{}{}.
\newblock
{\BBOQ}\APACrefatitle {Refixation frequency and memory mechanisms in visual
  search} {Refixation frequency and memory mechanisms in visual search}.{\BBCQ}
\newblock
\APACjournalVolNumPages{Current Biology}{10}{19}{1209--1212}.
\PrintBackRefs{\CurrentBib}

\bibitem [\protect \citeauthoryear {%
Guo%
, Meints%
, Hall%
, Hall%
\BCBL {}\ \BBA {} Mills%
}{%
Guo%
\ \protect \BOthers {.}}{%
{\protect \APACyear {2009}}%
}]{%
guo2009left}
\APACinsertmetastar {%
guo2009left}%
\begin{APACrefauthors}%
Guo, K.%
, Meints, K.%
, Hall, C.%
, Hall, S.%
\BCBL {}\ \BBA {} Mills, D.%
\end{APACrefauthors}%
\unskip\
\newblock
\APACrefYearMonthDay{2009}{}{}.
\newblock
{\BBOQ}\APACrefatitle {Left gaze bias in humans, rhesus monkeys and domestic
  dogs} {Left gaze bias in humans, rhesus monkeys and domestic dogs}.{\BBCQ}
\newblock
\APACjournalVolNumPages{Animal cognition}{12}{3}{409--418}.
\PrintBackRefs{\CurrentBib}

\bibitem [\protect \citeauthoryear {%
Harel%
, Koch%
\BCBL {}\ \BBA {} Perona%
}{%
Harel%
\ \protect \BOthers {.}}{%
{\protect \APACyear {2006}}%
}]{%
harel2006graph}
\APACinsertmetastar {%
harel2006graph}%
\begin{APACrefauthors}%
Harel, J.%
, Koch, C.%
\BCBL {}\ \BBA {} Perona, P.%
\end{APACrefauthors}%
\unskip\
\newblock
\APACrefYearMonthDay{2006}{}{}.
\newblock
{\BBOQ}\APACrefatitle {Graph-based visual saliency} {Graph-based visual
  saliency}.{\BBCQ}
\newblock
\BIn{} \APACrefbtitle {Advances in neural information processing systems}
  {Advances in neural information processing systems}\ (\BPGS\ 545--552).
\PrintBackRefs{\CurrentBib}

\bibitem [\protect \citeauthoryear {%
Henderson%
}{%
Henderson%
}{%
{\protect \APACyear {2003}}%
}]{%
henderson2003human}
\APACinsertmetastar {%
henderson2003human}%
\begin{APACrefauthors}%
Henderson, J\BPBI M.%
\end{APACrefauthors}%
\unskip\
\newblock
\APACrefYearMonthDay{2003}{}{}.
\newblock
{\BBOQ}\APACrefatitle {Human gaze control during real-world scene perception}
  {Human gaze control during real-world scene perception}.{\BBCQ}
\newblock
\APACjournalVolNumPages{Trends in cognitive sciences}{7}{11}{498--504}.
\PrintBackRefs{\CurrentBib}

\bibitem [\protect \citeauthoryear {%
Henderson%
\ \BBA {} Hollingworth%
}{%
Henderson%
\ \BBA {} Hollingworth%
}{%
{\protect \APACyear {1998}}%
}]{%
henderson1998eye}
\APACinsertmetastar {%
henderson1998eye}%
\begin{APACrefauthors}%
Henderson, J\BPBI M.%
\BCBT {}\ \BBA {} Hollingworth, A.%
\end{APACrefauthors}%
\unskip\
\newblock
\APACrefYearMonthDay{1998}{}{}.
\newblock
{\BBOQ}\APACrefatitle {Eye movements during scene viewing: An overview} {Eye
  movements during scene viewing: An overview}.{\BBCQ}
\newblock
\APACjournalVolNumPages{Eye guidance in reading and scene
  perception}{11}{}{269--293}.
\PrintBackRefs{\CurrentBib}

\bibitem [\protect \citeauthoryear {%
Henderson%
\ \BBA {} Hollingworth%
}{%
Henderson%
\ \BBA {} Hollingworth%
}{%
{\protect \APACyear {2003}}%
}]{%
henderson2003eye}
\APACinsertmetastar {%
henderson2003eye}%
\begin{APACrefauthors}%
Henderson, J\BPBI M.%
\BCBT {}\ \BBA {} Hollingworth, A.%
\end{APACrefauthors}%
\unskip\
\newblock
\APACrefYearMonthDay{2003}{}{}.
\newblock
{\BBOQ}\APACrefatitle {Eye movements, visual memory, and scene representation}
  {Eye movements, visual memory, and scene representation}.{\BBCQ}
\newblock
\APACjournalVolNumPages{Perception of faces, objects, and
  scenes}{}{}{356--383}.
\PrintBackRefs{\CurrentBib}

\bibitem [\protect \citeauthoryear {%
Hooge%
, Over%
, van Wezel%
\BCBL {}\ \BBA {} Frens%
}{%
Hooge%
\ \protect \BOthers {.}}{%
{\protect \APACyear {2005}}%
}]{%
hooge2005inhibition}
\APACinsertmetastar {%
hooge2005inhibition}%
\begin{APACrefauthors}%
Hooge, I\BPBI T\BPBI C.%
, Over, E\BPBI A.%
, van Wezel, R\BPBI J.%
\BCBL {}\ \BBA {} Frens, M\BPBI A.%
\end{APACrefauthors}%
\unskip\
\newblock
\APACrefYearMonthDay{2005}{}{}.
\newblock
{\BBOQ}\APACrefatitle {Inhibition of return is not a foraging facilitator in
  saccadic search and free viewing} {Inhibition of return is not a foraging
  facilitator in saccadic search and free viewing}.{\BBCQ}
\newblock
\APACjournalVolNumPages{Vision research}{45}{14}{1901--1908}.
\PrintBackRefs{\CurrentBib}

\bibitem [\protect \citeauthoryear {%
Itti%
\ \BBA {} Koch%
}{%
Itti%
\ \BBA {} Koch%
}{%
{\protect \APACyear {2000}}%
}]{%
itti2000saliency}
\APACinsertmetastar {%
itti2000saliency}%
\begin{APACrefauthors}%
Itti, L.%
\BCBT {}\ \BBA {} Koch, C.%
\end{APACrefauthors}%
\unskip\
\newblock
\APACrefYearMonthDay{2000}{}{}.
\newblock
{\BBOQ}\APACrefatitle {A saliency-based search mechanism for overt and covert
  shifts of visual attention} {A saliency-based search mechanism for overt and
  covert shifts of visual attention}.{\BBCQ}
\newblock
\APACjournalVolNumPages{Vision research}{40}{10}{1489--1506}.
\PrintBackRefs{\CurrentBib}

\bibitem [\protect \citeauthoryear {%
Itti%
\ \BBA {} Koch%
}{%
Itti%
\ \BBA {} Koch%
}{%
{\protect \APACyear {2001}}%
}]{%
itti2001computational}
\APACinsertmetastar {%
itti2001computational}%
\begin{APACrefauthors}%
Itti, L.%
\BCBT {}\ \BBA {} Koch, C.%
\end{APACrefauthors}%
\unskip\
\newblock
\APACrefYearMonthDay{2001}{}{}.
\newblock
{\BBOQ}\APACrefatitle {Computational modelling of visual attention}
  {Computational modelling of visual attention}.{\BBCQ}
\newblock
\APACjournalVolNumPages{Nature reviews neuroscience}{2}{3}{194--203}.
\PrintBackRefs{\CurrentBib}

\bibitem [\protect \citeauthoryear {%
Itti%
, Koch%
\BCBL {}\ \BBA {} Niebur%
}{%
Itti%
\ \protect \BOthers {.}}{%
{\protect \APACyear {1998}}%
}]{%
itti1998model}
\APACinsertmetastar {%
itti1998model}%
\begin{APACrefauthors}%
Itti, L.%
, Koch, C.%
\BCBL {}\ \BBA {} Niebur, E.%
\end{APACrefauthors}%
\unskip\
\newblock
\APACrefYearMonthDay{1998}{}{}.
\newblock
{\BBOQ}\APACrefatitle {A model of saliency-based visual attention for rapid
  scene analysis} {A model of saliency-based visual attention for rapid scene
  analysis}.{\BBCQ}
\newblock
\APACjournalVolNumPages{IEEE Transactions on pattern analysis and machine
  intelligence}{20}{11}{1254--1259}.
\PrintBackRefs{\CurrentBib}

\bibitem [\protect \citeauthoryear {%
Judd%
, Durand%
\BCBL {}\ \BBA {} Torralba%
}{%
Judd%
\ \protect \BOthers {.}}{%
{\protect \APACyear {2012}}%
}]{%
judd2012benchmark}
\APACinsertmetastar {%
judd2012benchmark}%
\begin{APACrefauthors}%
Judd, T.%
, Durand, F.%
\BCBL {}\ \BBA {} Torralba, A.%
\end{APACrefauthors}%
\unskip\
\newblock
\APACrefYearMonthDay{2012}{}{}.
\newblock
{\BBOQ}\APACrefatitle {A benchmark of computational models of saliency to
  predict human fixations} {A benchmark of computational models of saliency to
  predict human fixations}.{\BBCQ}
\newblock
\APACjournalVolNumPages{Tech.Rep.TR, MIT-CSAIL}{2012-001}{}{}.
\PrintBackRefs{\CurrentBib}

\bibitem [\protect \citeauthoryear {%
Judd%
, Ehinger%
, Durand%
\BCBL {}\ \BBA {} Torralba%
}{%
Judd%
\ \protect \BOthers {.}}{%
{\protect \APACyear {2009}}%
}]{%
judd2009learning}
\APACinsertmetastar {%
judd2009learning}%
\begin{APACrefauthors}%
Judd, T.%
, Ehinger, K.%
, Durand, F.%
\BCBL {}\ \BBA {} Torralba, A.%
\end{APACrefauthors}%
\unskip\
\newblock
\APACrefYearMonthDay{2009}{}{}.
\newblock
{\BBOQ}\APACrefatitle {Learning to predict where humans look} {Learning to
  predict where humans look}.{\BBCQ}
\newblock
\BIn{} \APACrefbtitle {Computer Vision, 2009 IEEE 12th international conference
  on} {Computer vision, 2009 ieee 12th international conference on}\ (\BPGS\
  2106--2113).
\PrintBackRefs{\CurrentBib}

\bibitem [\protect \citeauthoryear {%
Kienzle%
, Wichmann%
, Franz%
\BCBL {}\ \BBA {} Sch{\"o}lkopf%
}{%
Kienzle%
\ \protect \BOthers {.}}{%
{\protect \APACyear {2006}}%
}]{%
kienzle2006nonparametric}
\APACinsertmetastar {%
kienzle2006nonparametric}%
\begin{APACrefauthors}%
Kienzle, W.%
, Wichmann, F\BPBI A.%
, Franz, M\BPBI O.%
\BCBL {}\ \BBA {} Sch{\"o}lkopf, B.%
\end{APACrefauthors}%
\unskip\
\newblock
\APACrefYearMonthDay{2006}{}{}.
\newblock
{\BBOQ}\APACrefatitle {A nonparametric approach to bottom-up visual saliency}
  {A nonparametric approach to bottom-up visual saliency}.{\BBCQ}
\newblock
\BIn{} \APACrefbtitle {Advances in neural information processing systems}
  {Advances in neural information processing systems}\ (\BPGS\ 689--696).
\PrintBackRefs{\CurrentBib}

\bibitem [\protect \citeauthoryear {%
Killian%
, Jutras%
\BCBL {}\ \BBA {} Buffalo%
}{%
Killian%
\ \protect \BOthers {.}}{%
{\protect \APACyear {2012}}%
}]{%
killian2012map}
\APACinsertmetastar {%
killian2012map}%
\begin{APACrefauthors}%
Killian, N\BPBI J.%
, Jutras, M\BPBI J.%
\BCBL {}\ \BBA {} Buffalo, E\BPBI A.%
\end{APACrefauthors}%
\unskip\
\newblock
\APACrefYearMonthDay{2012}{}{}.
\newblock
{\BBOQ}\APACrefatitle {A map of visual space in the primate entorhinal cortex}
  {A map of visual space in the primate entorhinal cortex}.{\BBCQ}
\newblock
\APACjournalVolNumPages{Nature}{491}{7426}{761--764}.
\PrintBackRefs{\CurrentBib}

\bibitem [\protect \citeauthoryear {%
Klein%
}{%
Klein%
}{%
{\protect \APACyear {1988}}%
}]{%
klein1988inhibitory}
\APACinsertmetastar {%
klein1988inhibitory}%
\begin{APACrefauthors}%
Klein, R.%
\end{APACrefauthors}%
\unskip\
\newblock
\APACrefYearMonthDay{1988}{}{}.
\newblock
{\BBOQ}\APACrefatitle {Inhibitory tagging system facilitates visual search}
  {Inhibitory tagging system facilitates visual search}.{\BBCQ}
\newblock
\APACjournalVolNumPages{Nature}{334}{6181}{430--431}.
\PrintBackRefs{\CurrentBib}

\bibitem [\protect \citeauthoryear {%
Klein%
}{%
Klein%
}{%
{\protect \APACyear {2000}}%
}]{%
klein2000inhibition}
\APACinsertmetastar {%
klein2000inhibition}%
\begin{APACrefauthors}%
Klein, R.%
\end{APACrefauthors}%
\unskip\
\newblock
\APACrefYearMonthDay{2000}{}{}.
\newblock
{\BBOQ}\APACrefatitle {Inhibition of return} {Inhibition of return}.{\BBCQ}
\newblock
\APACjournalVolNumPages{Trends in cognitive sciences}{4}{4}{138--147}.
\PrintBackRefs{\CurrentBib}

\bibitem [\protect \citeauthoryear {%
Kuznetsova%
, Brockhoff%
\BCBL {}\ \BBA {} Christensen%
}{%
Kuznetsova%
\ \protect \BOthers {.}}{%
{\protect \APACyear {2013}}%
}]{%
kuznetsova2013lmertest}
\APACinsertmetastar {%
kuznetsova2013lmertest}%
\begin{APACrefauthors}%
Kuznetsova, A.%
, Brockhoff, P\BPBI B.%
\BCBL {}\ \BBA {} Christensen, R\BPBI H\BPBI B.%
\end{APACrefauthors}%
\unskip\
\newblock
\APACrefYearMonthDay{2013}{}{}.
\newblock
{\BBOQ}\APACrefatitle {lmerTest: Tests for random and fixed effects for linear
  mixed effect models (lmer objects of lme4 package)} {lmertest: Tests for
  random and fixed effects for linear mixed effect models (lmer objects of lme4
  package)}.{\BBCQ}
\newblock
\APACjournalVolNumPages{R package version}{2}{6}{}.
\PrintBackRefs{\CurrentBib}

\bibitem [\protect \citeauthoryear {%
Le~Meur%
\ \BBA {} Liu%
}{%
Le~Meur%
\ \BBA {} Liu%
}{%
{\protect \APACyear {2015}}%
}]{%
le2015saccadic}
\APACinsertmetastar {%
le2015saccadic}%
\begin{APACrefauthors}%
Le~Meur, O.%
\BCBT {}\ \BBA {} Liu, Z.%
\end{APACrefauthors}%
\unskip\
\newblock
\APACrefYearMonthDay{2015}{}{}.
\newblock
{\BBOQ}\APACrefatitle {Saccadic model of eye movements for free-viewing
  condition} {Saccadic model of eye movements for free-viewing
  condition}.{\BBCQ}
\newblock
\APACjournalVolNumPages{Vision research}{}{}{}.
\PrintBackRefs{\CurrentBib}

\bibitem [\protect \citeauthoryear {%
Loftus%
\ \BBA {} Masson%
}{%
Loftus%
\ \BBA {} Masson%
}{%
{\protect \APACyear {1994}}%
}]{%
loftus1994using}
\APACinsertmetastar {%
loftus1994using}%
\begin{APACrefauthors}%
Loftus, G\BPBI R.%
\BCBT {}\ \BBA {} Masson, M\BPBI E.%
\end{APACrefauthors}%
\unskip\
\newblock
\APACrefYearMonthDay{1994}{}{}.
\newblock
{\BBOQ}\APACrefatitle {Using confidence intervals in within-subject designs}
  {Using confidence intervals in within-subject designs}.{\BBCQ}
\newblock
\APACjournalVolNumPages{Psychonomic bulletin \& review}{1}{4}{476--490}.
\PrintBackRefs{\CurrentBib}

\bibitem [\protect \citeauthoryear {%
Luce%
}{%
Luce%
}{%
{\protect \APACyear {1959}}%
}]{%
luce1959individual}
\APACinsertmetastar {%
luce1959individual}%
\begin{APACrefauthors}%
Luce, R\BPBI D.%
\end{APACrefauthors}%
\unskip\
\newblock
\APACrefYear{1959}.
\newblock
\APACrefbtitle {Individual choice behavior: A theoretical analysis} {Individual
  choice behavior: A theoretical analysis}.
\newblock
\APACaddressPublisher{}{New York: Wiley}.
\PrintBackRefs{\CurrentBib}

\bibitem [\protect \citeauthoryear {%
Luke%
, Schmidt%
\BCBL {}\ \BBA {} Henderson%
}{%
Luke%
\ \protect \BOthers {.}}{%
{\protect \APACyear {2013}}%
}]{%
luke2013temporal}
\APACinsertmetastar {%
luke2013temporal}%
\begin{APACrefauthors}%
Luke, S\BPBI G.%
, Schmidt, J.%
\BCBL {}\ \BBA {} Henderson, J\BPBI M.%
\end{APACrefauthors}%
\unskip\
\newblock
\APACrefYearMonthDay{2013}{}{}.
\newblock
{\BBOQ}\APACrefatitle {Temporal oculomotor inhibition of return and spatial
  facilitation of return in a visual encoding task} {Temporal oculomotor
  inhibition of return and spatial facilitation of return in a visual encoding
  task}.{\BBCQ}
\newblock
\APACjournalVolNumPages{Frontiers in psychology}{4}{}{}.
\PrintBackRefs{\CurrentBib}

\bibitem [\protect \citeauthoryear {%
Mannan%
, Ruddock%
\BCBL {}\ \BBA {} Wooding%
}{%
Mannan%
\ \protect \BOthers {.}}{%
{\protect \APACyear {1996}}%
}]{%
mannan1996relationship}
\APACinsertmetastar {%
mannan1996relationship}%
\begin{APACrefauthors}%
Mannan, S\BPBI K.%
, Ruddock, K\BPBI H.%
\BCBL {}\ \BBA {} Wooding, D\BPBI S.%
\end{APACrefauthors}%
\unskip\
\newblock
\APACrefYearMonthDay{1996}{}{}.
\newblock
{\BBOQ}\APACrefatitle {The relationship between the locations of spatial
  features and those of fixations made during visual examination of briefly
  presented images} {The relationship between the locations of spatial features
  and those of fixations made during visual examination of briefly presented
  images}.{\BBCQ}
\newblock
\APACjournalVolNumPages{Spatial Vision}{10}{3}{165--188}.
\PrintBackRefs{\CurrentBib}

\bibitem [\protect \citeauthoryear {%
Navalpakkam%
\ \BBA {} Itti%
}{%
Navalpakkam%
\ \BBA {} Itti%
}{%
{\protect \APACyear {2005}}%
}]{%
navalpakkam2005modeling}
\APACinsertmetastar {%
navalpakkam2005modeling}%
\begin{APACrefauthors}%
Navalpakkam, V.%
\BCBT {}\ \BBA {} Itti, L.%
\end{APACrefauthors}%
\unskip\
\newblock
\APACrefYearMonthDay{2005}{}{}.
\newblock
{\BBOQ}\APACrefatitle {Modeling the influence of task on attention} {Modeling
  the influence of task on attention}.{\BBCQ}
\newblock
\APACjournalVolNumPages{Vision research}{45}{2}{205--231}.
\PrintBackRefs{\CurrentBib}

\bibitem [\protect \citeauthoryear {%
Ossand{\'o}n%
, Onat%
\BCBL {}\ \BBA {} K{\"o}nig%
}{%
Ossand{\'o}n%
\ \protect \BOthers {.}}{%
{\protect \APACyear {2014}}%
}]{%
ossandon2014spatial}
\APACinsertmetastar {%
ossandon2014spatial}%
\begin{APACrefauthors}%
Ossand{\'o}n, J\BPBI P.%
, Onat, S.%
\BCBL {}\ \BBA {} K{\"o}nig, P.%
\end{APACrefauthors}%
\unskip\
\newblock
\APACrefYearMonthDay{2014}{}{}.
\newblock
{\BBOQ}\APACrefatitle {Spatial biases in viewing behavior} {Spatial biases in
  viewing behavior}.{\BBCQ}
\newblock
\APACjournalVolNumPages{Journal of vision}{14}{2}{20}.
\PrintBackRefs{\CurrentBib}

\bibitem [\protect \citeauthoryear {%
Parkhurst%
, Law%
\BCBL {}\ \BBA {} Niebur%
}{%
Parkhurst%
\ \protect \BOthers {.}}{%
{\protect \APACyear {2002}}%
}]{%
parkhurst2002modeling}
\APACinsertmetastar {%
parkhurst2002modeling}%
\begin{APACrefauthors}%
Parkhurst, D.%
, Law, K.%
\BCBL {}\ \BBA {} Niebur, E.%
\end{APACrefauthors}%
\unskip\
\newblock
\APACrefYearMonthDay{2002}{}{}.
\newblock
{\BBOQ}\APACrefatitle {Modeling the role of salience in the allocation of overt
  visual attention} {Modeling the role of salience in the allocation of overt
  visual attention}.{\BBCQ}
\newblock
\APACjournalVolNumPages{Vision research}{42}{1}{107--123}.
\PrintBackRefs{\CurrentBib}

\bibitem [\protect \citeauthoryear {%
Posner%
\ \BBA {} Cohen%
}{%
Posner%
\ \BBA {} Cohen%
}{%
{\protect \APACyear {1984}}%
}]{%
posner1984components}
\APACinsertmetastar {%
posner1984components}%
\begin{APACrefauthors}%
Posner, M\BPBI I.%
\BCBT {}\ \BBA {} Cohen, Y.%
\end{APACrefauthors}%
\unskip\
\newblock
\APACrefYearMonthDay{1984}{}{}.
\newblock
{\BBOQ}\APACrefatitle {Components of visual orienting} {Components of visual
  orienting}.{\BBCQ}
\newblock
\APACjournalVolNumPages{Attention and performance X: Control of language
  processes}{32}{}{531--556}.
\PrintBackRefs{\CurrentBib}

\bibitem [\protect \citeauthoryear {%
Posner%
, Rafal%
, Choate%
\BCBL {}\ \BBA {} Vaughan%
}{%
Posner%
\ \protect \BOthers {.}}{%
{\protect \APACyear {1985}}%
}]{%
posner1985inhibition}
\APACinsertmetastar {%
posner1985inhibition}%
\begin{APACrefauthors}%
Posner, M\BPBI I.%
, Rafal, R\BPBI D.%
, Choate, L\BPBI S.%
\BCBL {}\ \BBA {} Vaughan, J.%
\end{APACrefauthors}%
\unskip\
\newblock
\APACrefYearMonthDay{1985}{}{}.
\newblock
{\BBOQ}\APACrefatitle {Inhibition of return: Neural basis and function}
  {Inhibition of return: Neural basis and function}.{\BBCQ}
\newblock
\APACjournalVolNumPages{Cognitive neuropsychology}{2}{3}{211--228}.
\PrintBackRefs{\CurrentBib}

\bibitem [\protect \citeauthoryear {%
{R Core Team}%
}{%
{R Core Team}%
}{%
{\protect \APACyear {2014}}%
}]{%
R}
\APACinsertmetastar {%
R}%
\begin{APACrefauthors}%
{R Core Team}.%
\end{APACrefauthors}%
\unskip\
\newblock
\APACrefYearMonthDay{2014}{}{}.
\newblock
{\BBOQ}\APACrefatitle {R: A Language and Environment for Statistical Computing}
  {R: A language and environment for statistical computing}{\BBCQ}\
  [\bibcomputersoftwaremanual].
\newblock
\APACaddressPublisher{Vienna, Austria}{}.
\newblock
\begin{APACrefURL} \url{http://www.R-project.org/} \end{APACrefURL}
\PrintBackRefs{\CurrentBib}

\bibitem [\protect \citeauthoryear {%
Satterthwaite%
}{%
Satterthwaite%
}{%
{\protect \APACyear {1946}}%
}]{%
satterthwaite1946approximate}
\APACinsertmetastar {%
satterthwaite1946approximate}%
\begin{APACrefauthors}%
Satterthwaite, F\BPBI E.%
\end{APACrefauthors}%
\unskip\
\newblock
\APACrefYearMonthDay{1946}{}{}.
\newblock
{\BBOQ}\APACrefatitle {An approximate distribution of estimates of variance
  components} {An approximate distribution of estimates of variance
  components}.{\BBCQ}
\newblock
\APACjournalVolNumPages{Biometrics bulletin}{}{}{110--114}.
\PrintBackRefs{\CurrentBib}

\bibitem [\protect \citeauthoryear {%
Scott%
}{%
Scott%
}{%
{\protect \APACyear {2012}}%
}]{%
scott2012multivariate}
\APACinsertmetastar {%
scott2012multivariate}%
\begin{APACrefauthors}%
Scott, D\BPBI W.%
\end{APACrefauthors}%
\unskip\
\newblock
\APACrefYearMonthDay{2012}{}{}.
\newblock
{\BBOQ}\APACrefatitle {Multivariate density estimation and visualization}
  {Multivariate density estimation and visualization}.{\BBCQ}
\newblock
\BIn{} \APACrefbtitle {Handbook of Computational Statistics} {Handbook of
  computational statistics}\ (\BPGS\ 549--569).
\newblock
\APACaddressPublisher{}{Springer}.
\PrintBackRefs{\CurrentBib}

\bibitem [\protect \citeauthoryear {%
Smith%
\ \BBA {} Henderson%
}{%
Smith%
\ \BBA {} Henderson%
}{%
{\protect \APACyear {2009}}%
}]{%
smith2009facilitation}
\APACinsertmetastar {%
smith2009facilitation}%
\begin{APACrefauthors}%
Smith, T\BPBI J.%
\BCBT {}\ \BBA {} Henderson, J\BPBI M.%
\end{APACrefauthors}%
\unskip\
\newblock
\APACrefYearMonthDay{2009}{}{}.
\newblock
{\BBOQ}\APACrefatitle {Facilitation of return during scene viewing}
  {Facilitation of return during scene viewing}.{\BBCQ}
\newblock
\APACjournalVolNumPages{Visual Cognition}{17}{6-7}{1083--1108}.
\PrintBackRefs{\CurrentBib}

\bibitem [\protect \citeauthoryear {%
Smith%
\ \BBA {} Henderson%
}{%
Smith%
\ \BBA {} Henderson%
}{%
{\protect \APACyear {2011}}%
}]{%
smith2011does}
\APACinsertmetastar {%
smith2011does}%
\begin{APACrefauthors}%
Smith, T\BPBI J.%
\BCBT {}\ \BBA {} Henderson, J\BPBI M.%
\end{APACrefauthors}%
\unskip\
\newblock
\APACrefYearMonthDay{2011}{}{}.
\newblock
{\BBOQ}\APACrefatitle {Does oculomotor inhibition of return influence fixation
  probability during scene search?} {Does oculomotor inhibition of return
  influence fixation probability during scene search?}{\BBCQ}
\newblock
\APACjournalVolNumPages{Attention, Perception, \&
  Psychophysics}{73}{8}{2384--2398}.
\PrintBackRefs{\CurrentBib}

\bibitem [\protect \citeauthoryear {%
Stensola%
\ \protect \BOthers {.}}{%
Stensola%
\ \protect \BOthers {.}}{%
{\protect \APACyear {2012}}%
}]{%
stensola2012entorhinal}
\APACinsertmetastar {%
stensola2012entorhinal}%
\begin{APACrefauthors}%
Stensola, H.%
, Stensola, T.%
, Solstad, T.%
, Fr{\o}land, K.%
, Moser, M\BHBI B.%
\BCBL {}\ \BBA {} Moser, E\BPBI I.%
\end{APACrefauthors}%
\unskip\
\newblock
\APACrefYearMonthDay{2012}{}{}.
\newblock
{\BBOQ}\APACrefatitle {The entorhinal grid map is discretized} {The entorhinal
  grid map is discretized}.{\BBCQ}
\newblock
\APACjournalVolNumPages{Nature}{492}{7427}{72--78}.
\PrintBackRefs{\CurrentBib}

\bibitem [\protect \citeauthoryear {%
Tatler%
}{%
Tatler%
}{%
{\protect \APACyear {2007}}%
}]{%
tatler2007central}
\APACinsertmetastar {%
tatler2007central}%
\begin{APACrefauthors}%
Tatler, B\BPBI W.%
\end{APACrefauthors}%
\unskip\
\newblock
\APACrefYearMonthDay{2007}{}{}.
\newblock
{\BBOQ}\APACrefatitle {The central fixation bias in scene viewing: Selecting an
  optimal viewing position independently of motor biases and image feature
  distributions} {The central fixation bias in scene viewing: Selecting an
  optimal viewing position independently of motor biases and image feature
  distributions}.{\BBCQ}
\newblock
\APACjournalVolNumPages{Journal of Vision}{7}{14}{4}.
\PrintBackRefs{\CurrentBib}

\bibitem [\protect \citeauthoryear {%
Tatler%
, Baddeley%
\BCBL {}\ \BBA {} Vincent%
}{%
Tatler%
\ \protect \BOthers {.}}{%
{\protect \APACyear {2006}}%
}]{%
tatler2006long}
\APACinsertmetastar {%
tatler2006long}%
\begin{APACrefauthors}%
Tatler, B\BPBI W.%
, Baddeley, R\BPBI J.%
\BCBL {}\ \BBA {} Vincent, B\BPBI T.%
\end{APACrefauthors}%
\unskip\
\newblock
\APACrefYearMonthDay{2006}{}{}.
\newblock
{\BBOQ}\APACrefatitle {The long and the short of it: Spatial statistics at
  fixation vary with saccade amplitude and task} {The long and the short of it:
  Spatial statistics at fixation vary with saccade amplitude and task}.{\BBCQ}
\newblock
\APACjournalVolNumPages{Vision research}{46}{12}{1857--1862}.
\PrintBackRefs{\CurrentBib}

\bibitem [\protect \citeauthoryear {%
Tatler%
\ \BBA {} Vincent%
}{%
Tatler%
\ \BBA {} Vincent%
}{%
{\protect \APACyear {2008}}%
}]{%
tatler2008systematic}
\APACinsertmetastar {%
tatler2008systematic}%
\begin{APACrefauthors}%
Tatler, B\BPBI W.%
\BCBT {}\ \BBA {} Vincent, B\BPBI T.%
\end{APACrefauthors}%
\unskip\
\newblock
\APACrefYearMonthDay{2008}{}{}.
\newblock
{\BBOQ}\APACrefatitle {Systematic tendencies in scene viewing} {Systematic
  tendencies in scene viewing}.{\BBCQ}
\newblock
\APACjournalVolNumPages{Journal of Eye Movement Research}{2}{2}{1--18}.
\PrintBackRefs{\CurrentBib}

\bibitem [\protect \citeauthoryear {%
Torralba%
}{%
Torralba%
}{%
{\protect \APACyear {2003}}%
}]{%
torralba2003modeling}
\APACinsertmetastar {%
torralba2003modeling}%
\begin{APACrefauthors}%
Torralba, A.%
\end{APACrefauthors}%
\unskip\
\newblock
\APACrefYearMonthDay{2003}{}{}.
\newblock
{\BBOQ}\APACrefatitle {Modeling global scene factors in attention} {Modeling
  global scene factors in attention}.{\BBCQ}
\newblock
\APACjournalVolNumPages{JOSA A}{20}{7}{1407--1418}.
\PrintBackRefs{\CurrentBib}

\bibitem [\protect \citeauthoryear {%
Wilming%
, Harst%
, Schmidt%
\BCBL {}\ \BBA {} K{\"o}nig%
}{%
Wilming%
\ \protect \BOthers {.}}{%
{\protect \APACyear {2013}}%
}]{%
wilming2013saccadic}
\APACinsertmetastar {%
wilming2013saccadic}%
\begin{APACrefauthors}%
Wilming, N.%
, Harst, S.%
, Schmidt, N.%
\BCBL {}\ \BBA {} K{\"o}nig, P.%
\end{APACrefauthors}%
\unskip\
\newblock
\APACrefYearMonthDay{2013}{}{}.
\newblock
{\BBOQ}\APACrefatitle {Saccadic momentum and facilitation of return saccades
  contribute to an optimal foraging strategy} {Saccadic momentum and
  facilitation of return saccades contribute to an optimal foraging
  strategy}.{\BBCQ}
\newblock
\APACjournalVolNumPages{PLoS computational biology}{9}{1}{e1002871}.
\PrintBackRefs{\CurrentBib}

\bibitem [\protect \citeauthoryear {%
Yarbus%
, Haigh%
\BCBL {}\ \BBA {} Rigss%
}{%
Yarbus%
\ \protect \BOthers {.}}{%
{\protect \APACyear {1967}}%
}]{%
yarbus1967eye}
\APACinsertmetastar {%
yarbus1967eye}%
\begin{APACrefauthors}%
Yarbus, A\BPBI L.%
, Haigh, B.%
\BCBL {}\ \BBA {} Rigss, L\BPBI A.%
\end{APACrefauthors}%
\unskip\
\newblock
\APACrefYear{1967}.
\newblock
\APACrefbtitle {Eye movements and vision} {Eye movements and vision}\
  (\BVOL~2)\ (\BNUM\ 5.10).
\newblock
\APACaddressPublisher{}{Plenum press New York}.
\PrintBackRefs{\CurrentBib}

\end{thebibliography}

\end{document}